\documentclass[12pt]{article}

\usepackage{graphicx}
\usepackage{amsmath,amssymb}
\usepackage{graphics}
\usepackage{graphicx}
\usepackage[latin1]{inputenc}
\usepackage{epsfig}

\newcommand{\R}{\mathbb{R}}

\newcommand{\E}{\mathbb{E}}

\newcommand{\pr}{\mathcal{P}}
\newcommand{\br}{\mathcal{B}}
\newcommand{\calH}{\mathcal{H}}

\newcommand{\mediaX}{ {\bar{\cal X}}}

\begin{document}

\date{}

\title{\vspace{-0.8cm}\kern-2.0truecm{\LARGE Systemic risk governance   in a dynamical model of a banking system}}


\author{\kern-0.0truecm
\normalsize{\bf Lorella Fatone} \\
{\kern-0.0truecm\small Dipartimento di Matematica}\\
{\kern-0.0truecm\small Universit\`a di Camerino}\\
{\kern-0.0truecm\small  Via Madonna delle Carceri 9, 62032
Camerino (MC),
Italy}\\
{\kern-0.0truecm\small Ph. n.+39-0737-402558, FAX
n.+39-0737-632525,
E-mail: lorella.fatone@unicam.it}\\[4mm]
\normalsize {\kern-0.0truecm\bf Francesca Mariani}\\
{\kern-0.0truecm\small   Dipartimento di Scienze Economiche e Sociali}\\
{\kern-0.0truecm\small  Universit\`a Politecnica delle Marche}\\
{\kern-0.0truecm\small  Piazza Martelli 8, 60121 Ancona (AN), Italy}\\
{\kern-0.0truecm\small Ph. n.+39-071-2207243, FAX
n.+39-071-2207102,
E-mail: f.mariani@univpm.it}\\[4mm]
 }
\maketitle
%
%
\begin{abstract}
\noindent
We consider  the problem of governing  systemic risk in  a    banking system model.  The banking system model consists in   an initial value problem for a system of stochastic differential equations whose  dependent variables are  the   log-monetary reserves of the banks   as  functions of time.
The banking system model  considered  generalizes previous   models  studied    in  \cite{Fouque1},  \cite{Fouque2},  \cite {Papan1} 
and describes an homogeneous population of banks.
Two distinct   mechanisms are used to model    the cooperation among  banks and the cooperation between  banks and  monetary authority. These mechanisms are regulated respectively  by  the parameters  $\alpha$ and $\gamma$. 
%
 A bank fails  when its   log-monetary reserves go below an assigned default level.  
 We call systemic  risk or systemic event  in a   bounded  time interval   the fact that  in that time interval  at least  a given fraction of the   banks   fails. 
The probability   of    systemic risk   in a  bounded time interval  is evaluated   using statistical simulation.  
A  method   to  govern    the probability of systemic risk  in a  bounded  time interval is presented. 
The  goal of the     governance  is to   keep  the  probability of systemic risk in a  bounded  time interval between  two  given thresholds.
The       governance  is  based on  the choice of  the  log-monetary reserves of  a kind of  ``ideal bank''  as a function of time     and  on  the solution of   an optimal control problem for the mean field approximation of the  banking system  model.
The solution of the optimal control   problem  determines  the parameters $\alpha$ and $\gamma$ as functions of time, that is defines the  rules of the   borrowing and lending activity among  banks and  between  banks and  monetary authority. 
Some numerical examples    are discussed.  In particular during a  two year period    we consider   the governance   of  systemic risk   in the next  year when governance  decisions are taken  quarterly. 
The     systemic risk governance  is tested  in  absence and  in presence of    positive and  negative shocks acting on the banking system. 
%
 %
%
\end{abstract}
\date
\maketitle
\section{Introduction}\label{sec1}
%
Given an interacting system made of agents that can fail individually, the  systemic risk associated to the system  is the risk that a large number  of agents fails simultaneously leading to the  failure of the entire system. 
Physical systems, epidemiological systems, engineering systems   and banking systems  are  some of the numerous  examples   where  systemic risk  occurs. Given the relevance of these   systems it is easy to understand that     assessment and   governance   of systemic risk  are  important issues.  
For example  in the case of banking systems it is well known that  in developed countries the regulation and the correct functioning  of  these   systems  are  necessary     for  the well being of the  whole economy. 
In these  countries  specific political and technical authorities    are responsible for the banking system management.    For simplicity we identify   these  authorities with the  monetary authority.

For a survey  of the multifaceted aspects of  systemic risk   we refer to \cite{HandSR},  \cite{SRbook} and to the references therein.
In particular in recent times   mathematical models  have been used to   study    systemic risk in the financial sector, see, for example,  \cite{Battiston},  \cite{Beale},  \cite{Fouque1},  \cite{Fouque2}, \cite{Haldane}, \cite{May}, \cite{Nier},  \cite{Papan1}.
In this paper we  are concerned with  assessment and  governance of systemic risk in  a mathematical model of a banking system.  The    agents of our  model  are the financial institutions, that, for simplicity, are identified  with the banks that  are represented   through their log-monetary reserves. The   log-monetary reserves  of a bank are simply the logarithm of the   monetary reserves of the bank. 

%

The banking system model  presented  generalizes  the models studied in   \cite{Fouque1},  \cite{Fouque2},  \cite{Papan1} and describes  an homogeneous population of banks. 
In the model  the log-monetary reserves of the banks as functions of time are represented as   interacting diffusion processes    defined implicitly by    a system of stochastic differential equations (see equations   (\ref{OurBanks12}),   (\ref{OurBanksCI12}), (\ref{rho})). 
The model   consists   in an initial value problem for this  system of stochastic differential equations.
In the model  each bank interacts  with the  other banks and with the monetary authority.
We say that  a bank is failed   when its   log-monetary reserves go   below  a  given level called  default  level and that   systemic risk in a bounded time interval  occurs when in  that   time interval  at least 
 a given fraction of the banks   of  the model  fails.
Two cooperation   mechanisms  act  in the model.  The first one is 
the  borrowing and lending  activity among  banks. The second one is the  borrowing and lending  activity between  banks and   monetary authority. These mechanisms   are  governed respectively by the  parameters  $\alpha$  and $\gamma$   (see equation   (\ref{OurBanks12})).  
The  mechanism that models the  borrowing and lending  activity among banks is  the same one used  in  \cite{Fouque1}. 
The  borrowing and lending  activity between  banks and   monetary authority depends from the difference as  function of time  between  the empirical  mean of the log-monetary reserves  of the  banks of the system and a  target   trajectory,  chosen by  the monetary authority,   that represents  the log-monetary reserves as a function of time  of  a kind of  ``ideal bank''. 
Note that since   in  the model the bank   population  is homogeneous    all the  banks  should behave as      the ``ideal bank''. 
The banking system  model  presented  in this paper  (i.e.  equations   (\ref{OurBanks12}),   (\ref{OurBanksCI12}), (\ref{rho}))  generalizes  the   model introduced  in  \cite{Fouque1}  where only the inter-bank  cooperation   mechanism  is considered and uses ideas taken from  \cite{Fouque2},  \cite{Papan1}.

 We study the properties of the  banking system  model  (i.e. of  equations      (\ref{OurBanks12}),   (\ref{OurBanksCI12}), (\ref{rho})), and  we  derive its  mean field approximation. Using statistical simulation we evaluate    the probability   of    systemic risk in a bounded time interval. 
We show  quantitatively  that  increasing the  rate of cooperation  among banks and between banks and  monetary authority increases the safety of  the individual bank, but,  in some circumstances,     increases   also the probability  of  ``extreme''  systemic risk in a 
bounded time interval. Extreme systemic risk means failure of all the banks or of almost all the banks.    
We present a  method  to govern  the probability of     systemic risk   in a 
bounded time interval based on    the  mean field  approximation   of the  banking system   model. 
 This method    pursues the goal of  keeping    the probability  of    systemic risk  in a bounded time interval between two given thresholds and   depends  from  the choice  made by the monetary authority    of  a    target trajectory  for  the     log-monetary reserves   of the ideal bank  as a function of time. This choice  is translated in rules for the banks (i.e. is translated in the  values    of  the    parameters $\alpha$ and $\gamma$  as  functions of time)   through the solution of  an  optimal control problem for the mean field approximation  of the  banking system  model.  

 We present  some numerical experiments     to  illustrate the behaviour of the banking system  model and of the systemic risk governance.  In particular we present the simulation of the systemic risk governance during a two year period. 
In the simulation governance decisions are taken  quarterly  and consist, at the beginning of each quarter,  in the evaluation of  
the  probability  of  systemic risk in the next  year   in absence of governance decisions,   in  the consequent choice   of  a   target trajectory for  the       log-monetary reserves   of the ideal bank  in  the next year and in the translation of this choice   in rules for the banks.
The  choice of the   target trajectory   of the log-monetary reserves  of the ideal bank      in the next year   is   done   with   a simple trial and error procedure.   The systemic risk governance  is tested in presence of positive and negative shocks  acting on the banking system.    

Note that   the    behaviour of    the banking  system model  is governed  through    its mean field approximation.  That is   a possibly high dimensional dynamical  model  (i.e. the banking system model)  is governed   with  a control law determined  using     a low dimensional dynamical model (i.e. the  mean field approximation  of the banking system model).  
This is  possible since  in an homogeneous  population of banks each bank behaves as a kind of mean bank and from the governance of  the mean bank it is possible to deduce how to govern the entire bank population.
 The same idea can be exploited     in     all the circumstances where an  homogeneous  population of agents is studied.

 
 The remainder of the paper is organized as follows.  In Section \ref{sec2}    the banking system model  developed 
 in   \cite{Fouque1}, \cite{Fouque2} is reviewed, moreover    failure of a bank and  systemic risk in a bounded time interval are defined. 
 In Section \ref{sec3}    the banking system models  studied  in this paper are    introduced,  their mean field approximation is deduced   and  the  probability   of systemic risk  in a  bounded time interval  is evaluated.  
In particular  in  Sections  \ref{sec2}  and  \ref{sec3}  we investigate,   in  the banking system models considered,   the relation between  safety  of the individual bank and safety   of the  banking  system. 
In Section \ref{sec4} we solve  an optimal control problem for the mean field approximations  of  the banking system   models studied  in  Section   \ref{sec3} and  we show  how to translate the solution of the   control problems studied   in rules  for the banks, that is  in    values    of   the parameters $\alpha$ and $\gamma$ as functions of time.   
  Finally in Section \ref{sec5}    a method   to govern  systemic risk in a bounded time interval  is  presented  and  tested  in some numerical examples.

 
%
\section{The Fouque and Sun banking  system model and the definition of  systemic risk in a bounded time interval }\label{sec2}
%
 
For the convenience of the reader  we  begin reviewing    the banking system models presented  in  \cite{Fouque1},  \cite{Fouque2}.
Let $t$ be a real variable that  denotes  time and  $N>1$ be the number of banks present in the banking system model  at time $t=0$.  
Note that due to the possible failure  of banks during the time evolution  the number of banks in the model is not necessarily constant. 
For  $i=1,2,\ldots,N$   let $Y_{t}^{i}$,  $ t>0$,  be a diffusion process representing the log-monetary reserves of the $i$-th bank as a function of time. 

 In  \cite{Fouque1}  the first  dynamical model   considered  to describe  a system of $N$ banks is   the following system of   stochastic differential equations:
\begin{eqnarray}
&& dY_{t}^{i}= \sigma dW_{t}^{i},\,  t>0, \quad i=1,2,\ldots,N,  \label{indBanks}
\end{eqnarray}
with the initial condition:
\begin{eqnarray}
&&Y_{0}^{i} =y_{0}^{i}, \quad i=1,2,\ldots,N,                  \label{indBanksCI}
\end{eqnarray}
where  $\sigma$ is a constant, the stochastic processes $W_{t}^{i},$ $ t>0$, $ i=1,2,\ldots,N,$   are standard Wiener processes such that $W_{0}^{i}=0,$ $ i=1,2,\ldots,N,$ and  $dW_{t}^{i},$   $t>0$,  $ i=1,2,\ldots,N,$ are their
stochastic differentials.  Moreover  as done  in  \cite{Fouque1}  we assume that:
\begin{eqnarray}\label{rho}
\E(dW_{t}^{i} dW_{t}^{j})&=&\delta_{i,j} dt,\,\, t>0, \, \,  i,j=1,2,\ldots,N,
\end{eqnarray}
where $\E(\cdot)$ denotes the expected value of $\cdot$
 and  $\delta_{i,j}$, $ i,j=1,2,\ldots,N$, is the Kronecker symbol.  Note that in  (\ref{indBanks})  the diffusion coefficient $\sigma$  is the same  in  all the equations. \\
The  initial conditions  $y_{0}^{i}$, $ i=1,2,\ldots,N,$   are random
variables that, for simplicity,    are  chosen   to be concentrated in a point with
probability one. We identify these random variables
 with the points where they are concentrated. 
 For  simplicity   we choose (see  \cite{Fouque1}):
 \begin{eqnarray}
&&Y_{0}^{i} =y_{0}^{i}=0, \quad i=1,2,\ldots,N.               \label{FBanksCI}
\end{eqnarray}
%

In  the model  (\ref{indBanks}), (\ref{FBanksCI}),  (\ref{rho}) all the banks are  equal,  that  is  the bank population described by   (\ref{indBanks}), (\ref{FBanksCI}),  (\ref{rho})   is homogeneous.  All the   banking system  models considered in this paper describe  homogeneous populations of banks.

In   \cite{Fouque1} Fouque and  Sun  introduce an interaction among  banks adding a set  of  drift terms  to  the 
model           (\ref{indBanks}), (\ref{FBanksCI}),  (\ref{rho}). These drift terms  represent  the rate of   inter-bank borrowing and lending activity. 
That is the  equations  (\ref{indBanks})  are  replaced  by   the  following system of  stochastic differential equations:
\begin{equation}\label{FBanks}
 dY_{t}^{i}= \frac{\alpha}{N}  \sum_{j=1}^{N} \left( Y_{t}^{j} -Y_{t}^{i}\right) dt+ \sigma dW_{t}^{i},\,  t>0, \quad i=1,2,\ldots,N ,  
\end{equation}
where $\alpha \ge 0$ is called  rate of mean-reversion.  The overall  rate of ``mean reversion''  $\displaystyle\frac{\alpha}{N} $ is normalized dividing  the parameter $\alpha$  by the number $N$ of banks,  see  \cite{Fouque1} for more details.

For  $i=1,2,\ldots,N$     the  cooperation  of  the $i$-th  bank  with the other banks   is described  by the term     $\displaystyle  \frac{\alpha}{N} \sum_{j=1}^{N} \left( Y_{t}^{j} -Y_{t}^{i}\right) dt$, $t>0$,  of  the  $i$-th equation of  system  (\ref{FBanks})   and is governed by  the parameter $\alpha$. 
These terms establish  that  for  $t>0$  and  for $ j=1,2,\ldots,N$,  if at time $t$   bank $j$ has more reserves than  bank $i$ (i.e. if  $ Y_{t}^{j} >Y_{t}^{i}$)  money flows from bank $j$ to  bank $i$, and this  flow is  proportional to the difference  $Y_{t}^{j} -Y_{t}^{i}$  at the  rate  $\displaystyle \frac{\alpha}{N}$,   the opposite happens if   bank $i$ has more reserves than  bank $j$
 (i.e. if  $ Y_{t}^{j} <Y_{t}^{i}$). The  interaction  mechanism  among  banks  given by    (\ref{FBanks})   is based on the idea that  ``who has more gives to those who have less'',   and     is a simple form of cooperative behaviour between banks. 
Note that when $\alpha=0$ there are no monetary flows among banks, and      system  (\ref{FBanks}) reduces  to   system  (\ref{indBanks}).

In  \cite{Fouque1}    the mean  field approximation  of the dynamical system (\ref{FBanks}), (\ref{FBanksCI}), (\ref{rho})  is studied.  In particular   it is shown that   the mean  field limit of     (\ref{FBanks}), (\ref{FBanksCI}), (\ref{rho})  when $N$   goes to infinity is the   initial value problem:
\begin{eqnarray}
&& d{\cal Y}_{t}= -\alpha {\cal Y}_{t}  dt+ \sigma dW_{t},\,  t>0, \label{FMF}\\
&& {\cal Y}_{0}= 0, \label{FMFCI}
\end{eqnarray}
where $W_{t}$, $ t>0$, is a standard Wiener process such that $W_{0}=0$  and  $dW_{t},$   $t>0$,  is its 
stochastic differential.   
The  stochastic process ${\cal Y}_{t}$, $t>0,$  solution of (\ref{FMF}), (\ref{FMFCI})  represents the time evolution of the log-monetary reserves of  a kind of    ``mean-bank''  of   the  bank  population  described by  (\ref{FBanks}), (\ref{FBanksCI}), (\ref{rho}). 
Due to the homogeneity of this bank population 
when $N$  goes to infinity all the banks of  model (\ref{FBanks}), (\ref{FBanksCI}), (\ref{rho})  behave  like  the ``mean-bank'' described by (\ref{FMF}), (\ref{FMFCI}). 

For a detailed explanation  of  mean field theory  in its original   context, that is in  the context of   statistical mechanics,   see, for example,  \cite{Gallavotti} and the references therein.

Given a   banking system model  (i.e.  equations (\ref{indBanks}), (\ref{FBanksCI}),  (\ref{rho}) or   (\ref{FBanks}), (\ref{FBanksCI}), (\ref{rho}) or ...) let us define the default of a bank in a bounded time interval  and its probability. 
For $i=1,2,\ldots,N$ the default of  the $i$-th  bank   in a  bounded time interval   is  defined as follows:  chosen   a default level   $D<0$  bank $i$ defaults  in a  bounded time interval  if  in that time interval its log-monetary reserves   $Y_{t}^{i}$, $t>0$,  reach   level $D$.  
%
%
That is, given $ 0\le \tau_{1}<\tau_{2} < + \infty$,  the time interval $[\tau_{1},\tau_{2}]$    and   a  default level  $D$,  we define the  event $F^{i}_{[\tau_{1},\tau_{2}]}$,     ``default   of the $i$-th bank in  the time interval $[\tau_{1},\tau_{2}]$'', as follows:
\begin{eqnarray}  \label{failure1}
&&  F^{i}_{[\tau_{1},\tau_{2}]}  =\left\{  \min_{\tau_{1}\le t\le \tau_{2}}  Y_{t}^{i} \le  D   \right\} , \quad i=1,2,\ldots,N. \quad   
\end{eqnarray}
%

Given a banking system model (for example  equations (\ref{indBanks}), (\ref{FBanksCI}),  (\ref{rho}) or   (\ref{FBanks}), (\ref{FBanksCI}), (\ref{rho}) or ...) for $ i=1,2,\ldots,N$ to the event ``default of the $i$-th bank in the time interval $[\tau_{1},\tau_{2}]$''  is associated a probability. 
Note that   we assume that   failed banks are removed from the   banking system  model considered. 
Let us give a  formal   definition  of  systemic risk  or  of systemic event   in a bounded time interval and  a quantitative measure  of  its  probability in a  banking system model. 
Let  $int\left[\cdot\right]$ be  the integer part of $\cdot$, and $M$ be a positive integer such that  $M \ge int\left[\frac{N}{2}\right]$, 
we define systemic  risk of type $M$,   or  systemic event  of type $M$,   in the  time interval $[\tau_{1},\tau_{2}]$ 
the event:  ``during  the time interval $[\tau_{1},\tau_{2}]$  at least $M$ banks fail''.  In this  paper we  choose $M = int\left[\frac{N}{2}\right]+1$.  That is  $SR_{[\tau_{1},\tau_{2}]}$, the systemic risk or  systemic event   of type $M=int\left[\frac{N}{2}\right]+1$ in  the time interval $[\tau_{1},\tau_{2}]$,  is  defined as follows:
\begin{equation}  \label{SR}
SR_{[\tau_{1},\tau_{2}]}\!\!=\!\!\left\{   \mbox{at least} \,  M \!= int\!\left[\frac{N}{2}\right] \!+\!1 \mbox{  banks fail in  the time interval } \, [\tau_{1},\tau_{2}]\right\} .   
\end{equation}

Let $\pr(\cdot)$ denote the probability of the event  $\cdot$.   Given a banking system model, using   statistical simulation,  we evaluate   the probability  of   the event    $F^{i}_{[\tau_{1},\tau_{2}]}$,      $i=1,2,\ldots,N, $  and    the probability of the event $SR_{[\tau_{1},\tau_{2}]}$.    
This is done   using  a set of numerically simulated trajectories of  the  banking system model considered and  approximating   the probability  of $F^{i}_{[\tau_{1},\tau_{2}]}$,      $i=1,2,\ldots,N, $ or  of $SR_{[\tau_{1},\tau_{2}]}$
 with the  corresponding frequency in the numerically simulated trajectories.

Note that the definition   (\ref{SR})  of systemic risk    in   the time interval $[\tau_{1},\tau_{2}]$    is  different from  the one used in \cite{Fouque1}. In  fact in \cite{Fouque1}    systemic  risk or systemic event  in the time interval $[\tau_{1},\tau_{2}]$ is  defined as follows:
\begin{eqnarray}  \label{failure2}
&&   \overline{SR}_{[\tau_{1},\tau_{2}]} = \left\{  \min_{\tau_{1}\le t\le \tau_{2}}   {\bar Y}_{t} \le D   \right\} , \quad   
\end{eqnarray}
where 
\begin{equation}\label{meanY}
 {\bar Y}_{t}= \frac{1}{N} \sum_{j=1}^{N} Y_{t}^{j} , \quad  t>0, 
\end{equation}
is the empirical mean of  log-monetary  reserves of the banks. 
 
 In  \cite{Fouque1} given  the default level $D<0$   the probability  $\pr ( \overline{SR}_{[\tau_{1},\tau_{2}]})$  of the event defined in  (\ref{failure2})  is studied  as a function of $\alpha$  when $N$ becomes large.
 The study of   $\pr ( \overline{SR}_{[\tau_{1},\tau_{2}]})$ is based on   the theory of large deviations.
 Note that the definition of failed bank  (\ref{failure1}) adopted here is the same one  adopted in \cite{Fouque1}. However  in \cite{Fouque1}      a    bank     failed     during the time evolution remains in the banking system model  and  continues to  be  part  of  the time evolution   defined by    (\ref{FBanks}), (\ref{FBanksCI}), (\ref{rho}).
This last  assumption makes easier  the   study  of  the probability 
$\pr ( \overline{SR}_{[\tau_{1},\tau_{2}]})$ using the  theory of large deviations, see  \cite{Fouque1}.  Instead we  assume that when  a bank  fails leaves  the banking system model. That is  after  defaulting the failed bank  does not continue to  be part of  the borrowing and lending activity  of  the banking system model  and the  variable that describes it is removed from the system of equations   (\ref{FBanks}).  The non failed banks continue their time evolution that is governed by  the dynamical model that remains after the removal. 
We prefer to use   (\ref{SR})  and the rules specified above  as  definition  of systemic risk in a time interval
 instead of  adopting the definition  and the rules given  in \cite{Fouque1}.  
In fact  we believe that the    assumptions made here   are   more realistic  than those  made in  \cite{Fouque1}. 
Moreover the probabilities  $\pr (SR_{[\tau_{1},\tau_{2}]})$  defined in   (\ref{SR})   and $\pr ( \overline{SR}_{[\tau_{1},\tau_{2}]})$    are both   easily evaluated  using statistical simulation.

 Fouque and  Sun in \cite{Fouque1}  show that the introduction of the terms $\displaystyle \frac{\alpha}{N}  \sum_{j=1}^{N} \left( Y_{t}^{j} -Y_{t}^{i}\right) dt$,  $t>0$,  $i=1,2,\ldots,N$,  in (\ref{FBanks}) stabilizes   the behaviour of  the individual bank (i.e. decreases the default probability of the individual bank) at the expense of increasing  the probability   of  extreme  systemic risk,   
 that is  the probability   of    systemic risk of type $M$ when $M$ is very close to $N$. 
In fact the trajectories  of     (\ref{FBanks}), (\ref{FBanksCI}), (\ref{rho})  when $\alpha>0$  are more grouped  near the straight line $y=0$ than  those of   (\ref{indBanks}), (\ref{FBanksCI}), (\ref{rho}).  This ``swarming'' or ``flocking'' effect of  the  trajectories  of    (\ref{FBanks}), (\ref{FBanksCI}), (\ref{rho})  is  a consequence of   the presence of the inter-bank  cooperation  mechanism   and   increases when $\alpha$ increases.  The same effect is present in the trajectories of the mean field  limit  initial value problem (\ref{FMF}), (\ref{FMFCI}) due to the  term $ -\alpha {\cal Y}_{t}$, $t>0$, ($\alpha>0$) in (\ref{FMF}).
That is  the  ``swarming''  trajectories  of   (\ref{FBanks}), (\ref{FBanksCI}), (\ref{rho})   reach the default level rarely (i.e.  the  stability of the individual bank increases when $\alpha$ increases), but  due to the    swarming  effect  when  they reach the default level they reach the default level all together and   systemic risk  occurs.
In particular the  probability of  extreme systemic risk   increases when $\alpha$ increases.




In Figures 1a)-3a) we show one trajectory  of  $Y_{t}^{1}$,  $ t \in [0,T]$, $T=1$,  solution of    (\ref{FBanks}), (\ref{FBanksCI}), (\ref{rho})  when $N=10$ and one trajectory  of $ {\cal Y}_{t}$, $ t \in [0,T]$, $T=1$,  solution of (\ref{FMF}), (\ref{FMFCI})  for three different values of $\alpha$, that is   for  $\alpha=1$ (Figure 1a)),  $\alpha=10$ (Figure 2a)) and $\alpha=100$ (Figure 3a)).  As  suggested  in \cite{Fouque1}  we choose  $D=-0.7$, $\sigma=1$ and we use the explicit Euler method   with time step $\Delta t=10^{-4} $ to integrate the initial value problems   (\ref{FBanks}), (\ref{FBanksCI}), (\ref{rho})  and (\ref{FMF}), (\ref{FMFCI})  from time $t=0$ up to time $t=T$, $T=1$.
Note that in Figures 1a)-3a)  it is sufficient to represent   one trajectory of  $Y_{t}^{1}$,  $ t \in [0,T]$, $T=1$,  solution of system   (\ref{FBanks}), (\ref{FBanksCI}), (\ref{rho}) to illustrate the behaviour of the corresponding  trajectory  of  system (\ref{FBanks}), (\ref{FBanksCI}), (\ref{rho}). In fact the bank population considered  is homogeneous and the processes   $Y_{t}^{i}$,  $ t \in [0,T]$, $T=1$, $i=1,2,\ldots,N$, behave all in the same way.
As expected, increasing  $\alpha$, that is  increasing the rate of cooperation   among banks,  increases the  ``swarming''  effect  around the line $y=0$ of the  trajectories  of   (\ref{FBanks}), (\ref{FBanksCI}), (\ref{rho}), and, as a consequence, decreases the number of defaults in  the time interval  $[0,T]$, $T=1$,   increasing  the stability of the system.
 In Figures  1a)-3a) 
the dotted line shows the default level $D=-0.7$.

Let us  consider  the loss distribution  in the  time interval  $[0,T]$,  $T=1$, i.e. the probability distribution  of the random variable: number of bank defaults   in the  time interval  $[0,T]$,  $T=1$. 
In  the model (\ref{FBanks}), (\ref{FBanksCI}), (\ref{rho})  when $N=10$ the loss distribution in  $[0,T]$,  $T=1$, is  evaluated   using statistical simulation    starting from  $10^{4}$ simulated trajectories of the model. 
Note that the loss distributions shown in Figures   1b)-3b)   are computed using the definition of systemic risk  (\ref{SR})  and using the assumptions   about bank failure made in this paper.
In Figures  1b)-3b) the dashed  line  shows the loss distribution in $[0,T]$,  $T=1$, of  system  (\ref{indBanks}), (\ref{FBanksCI}),  (\ref{rho})  when $N=10$,  while the solid line denotes the loss distribution of system  (\ref{FBanks}), (\ref{FBanksCI}), (\ref{rho})  when $N=10$ and  $\alpha=1$ (see Figure 1b)),  $\alpha=10$ (see Figure 2b)) and $\alpha=100$ (see Figure 3b)). 
Note that in the  case of independent diffusion processes, i.e.  when  we consider  system (\ref{indBanks}), (\ref{FBanksCI}),  (\ref{rho}),  or   when we consider  $\alpha=0$ in (\ref{FBanks}), (\ref{FBanksCI}), (\ref{rho}), the loss distribution  in  $[0,T]$,  $T=1$,  is a  binomial distribution with known parameters. 
%
 When $N=10$ let us look more closely to  the loss distribution  in  $[0,T]$,  $T=1$,   of the  diffusion processes solution of    (\ref{FBanks}), (\ref{FBanksCI}), (\ref{rho})   as a function of $\alpha$. This distribution  is  plotted  with a solid line  in Figures  1b)-3b). 
These distributions  have a large mass near zero defaults together with  a small, but not negligible, mass near $N$ defaults.  
These features of the loss distribution of (\ref{FBanks}), (\ref{FBanksCI}), (\ref{rho})   and their  behaviour as functions of  
$\alpha$ when $\alpha$ increases show  that  the cooperation  mechanism  among banks  present   in  (\ref{FBanks}), (\ref{FBanksCI}), (\ref{rho}) increases  the stability of the individual bank, however it    increases also the  probability of ``extreme'' systemic risk  in the time interval  $[0,T]$, $T=1$.  
This last fact is shown  in  Figure 4 where we zoom  the right-bottom corner of Figure 1b)-3b). Figure 4 shows  that  the ``tail'' of the  loss distribution raises when $\alpha$  increases. 

Note that the  different definitions of systemic risk used in \cite{Fouque1}  (i.e.   (\ref{failure2})) and in this paper (i.e.   (\ref{SR}))  do not change  the qualitative behaviour  of the  systemic risk probability as a function of     $\alpha$  shown   in  Figures 1-4.

 \begin{figure}
 \centerline{\includegraphics[height=6cm]{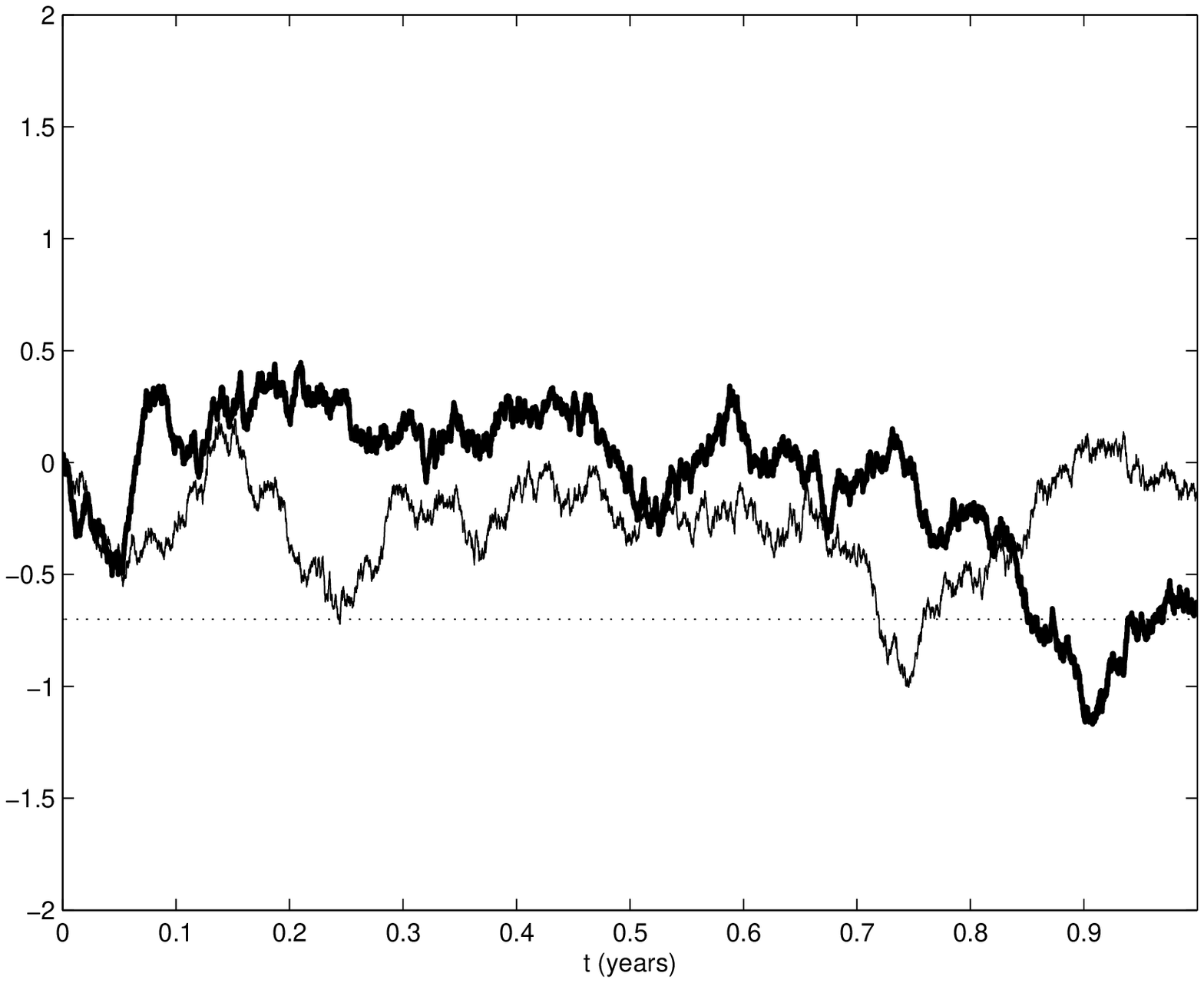}a)\includegraphics[height=6cm]{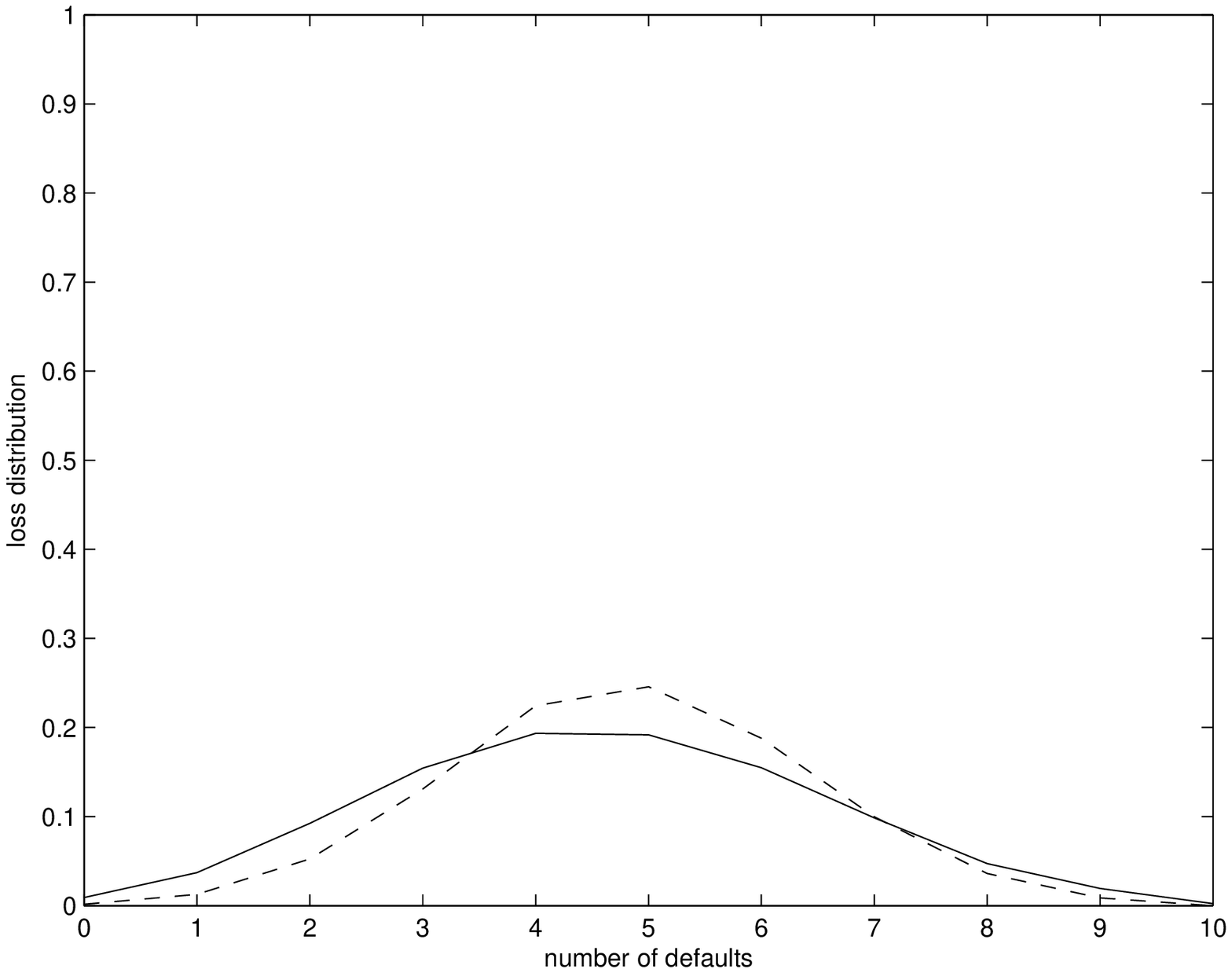}b)}
 \caption{\footnotesize a) 
 The solid line shows  one trajectory  of the  solution $Y_{t}^{1}$,  $ t \in [0,T]$, $T=1$,   of  system  (\ref{FBanks}), (\ref{FBanksCI}), (\ref{rho})   when   $N=10$, $\alpha=1$. 
 The thick solid line   shows  one trajectory  of the solution ${\cal Y}_{t}$,  $ t \in [0,T]$, $T=1$,   of  the  corresponding mean field initial value problem   (\ref{FMF}), (\ref{FMFCI}).
  The dotted  line shows the default level $D=-0.7$. 
 b)  Loss distribution in  $[0,T]$,  $T=1$,   of system  (\ref{FBanks}), (\ref{FBanksCI}), (\ref{rho}) when  $N=10$,   $\alpha=1$ (solid line) and  loss distribution in  $[0,T]$,  $T=1$,  of   system  (\ref{indBanks}), (\ref{FBanksCI}),  (\ref{rho}) when $N=10$ (dashed line).}
 \label{fig1}
  \end{figure}
 \begin{figure}
 \centerline{\includegraphics[height=6cm]{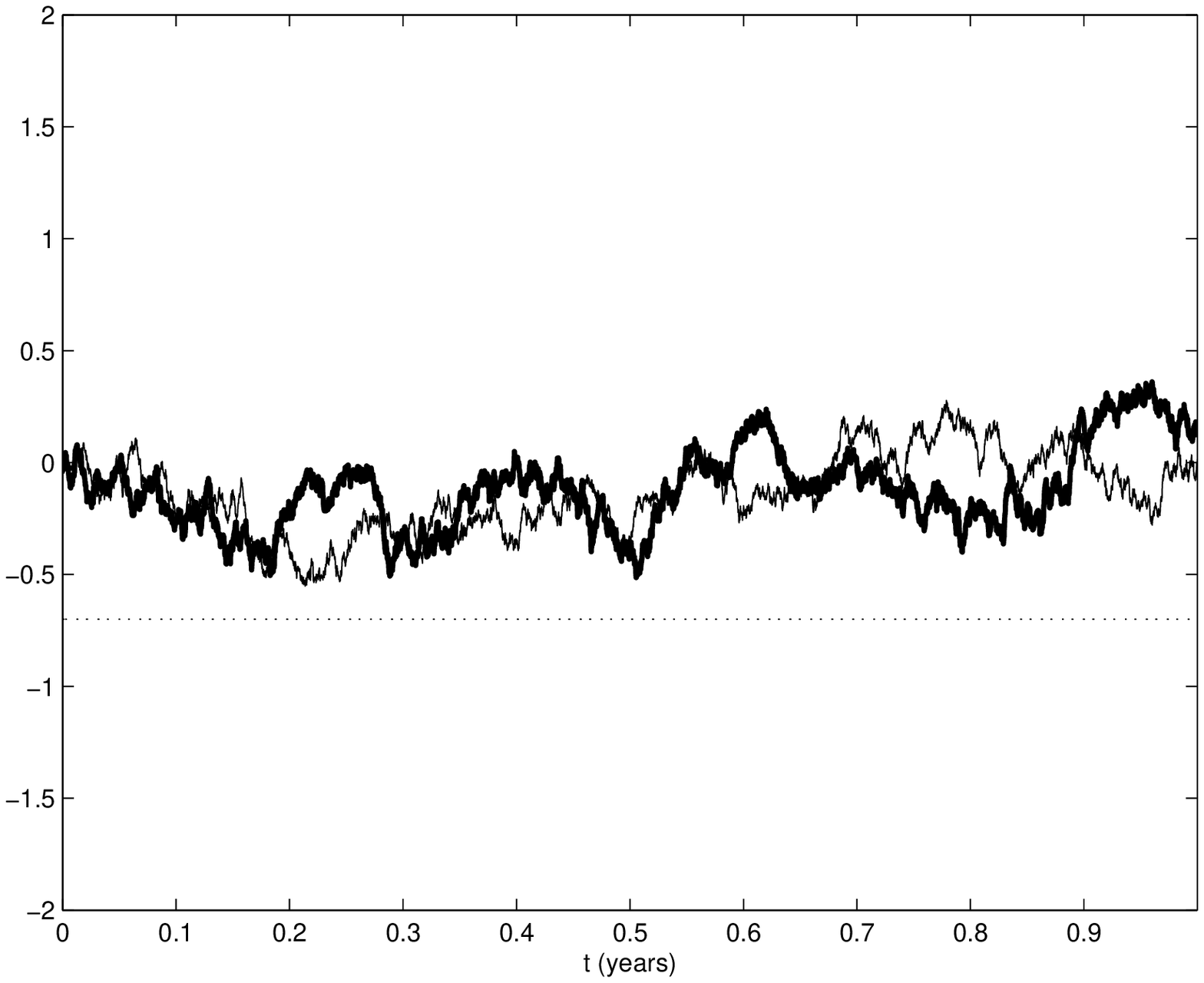}a)\includegraphics[height=6cm]{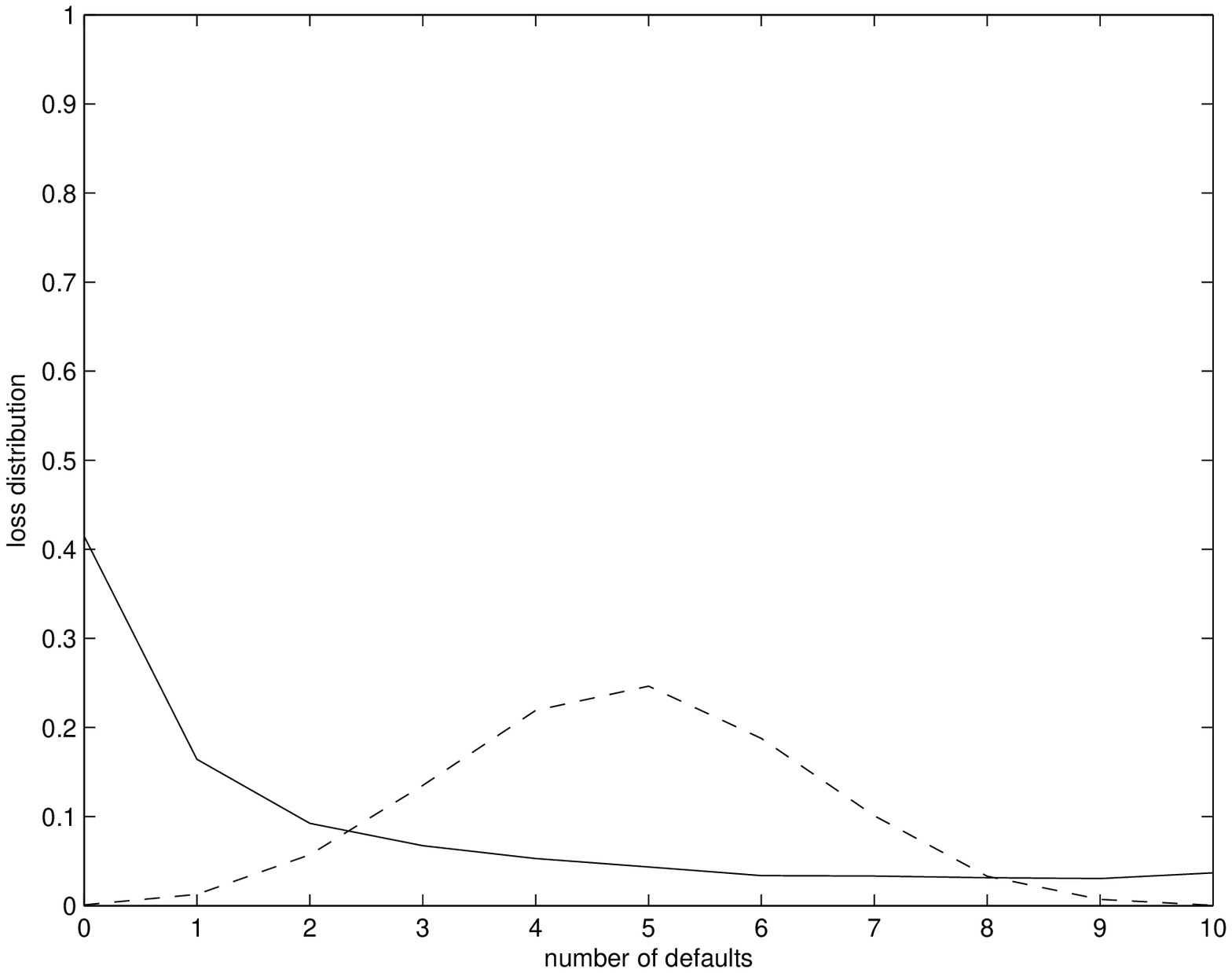}b)}
 \caption{\footnotesize a) 
 The solid line shows  one trajectory  of the solution $Y_{t}^{1}$,  $ t \in [0,T]$, $T=1$,  of  system  (\ref{FBanks}), (\ref{FBanksCI}), (\ref{rho})      when  
  $N=10$, $\alpha=10$. 
 The thick solid line   shows  one trajectory  of the  solution ${\cal Y}_{t}$,  $ t \in [0,T]$, $T=1$,  of  the  corresponding mean field initial value problem   (\ref{FMF}), (\ref{FMFCI}).
  The dotted  line shows the default level $D=-0.7$. 
 b)  Loss distribution in  $[0,T]$,  $T=1$,   of system  (\ref{FBanks}), (\ref{FBanksCI}), (\ref{rho}) when   $N=10$, $\alpha=10$ (solid line) and  loss distribution in  $[0,T]$,  $T=1$,  of   system  (\ref{indBanks}), (\ref{FBanksCI}),  (\ref{rho}) when $N=10$ (dashed line).}
 \label{fig2}
 \end{figure}
 \begin{figure}
 \centerline{\includegraphics[height=6cm]{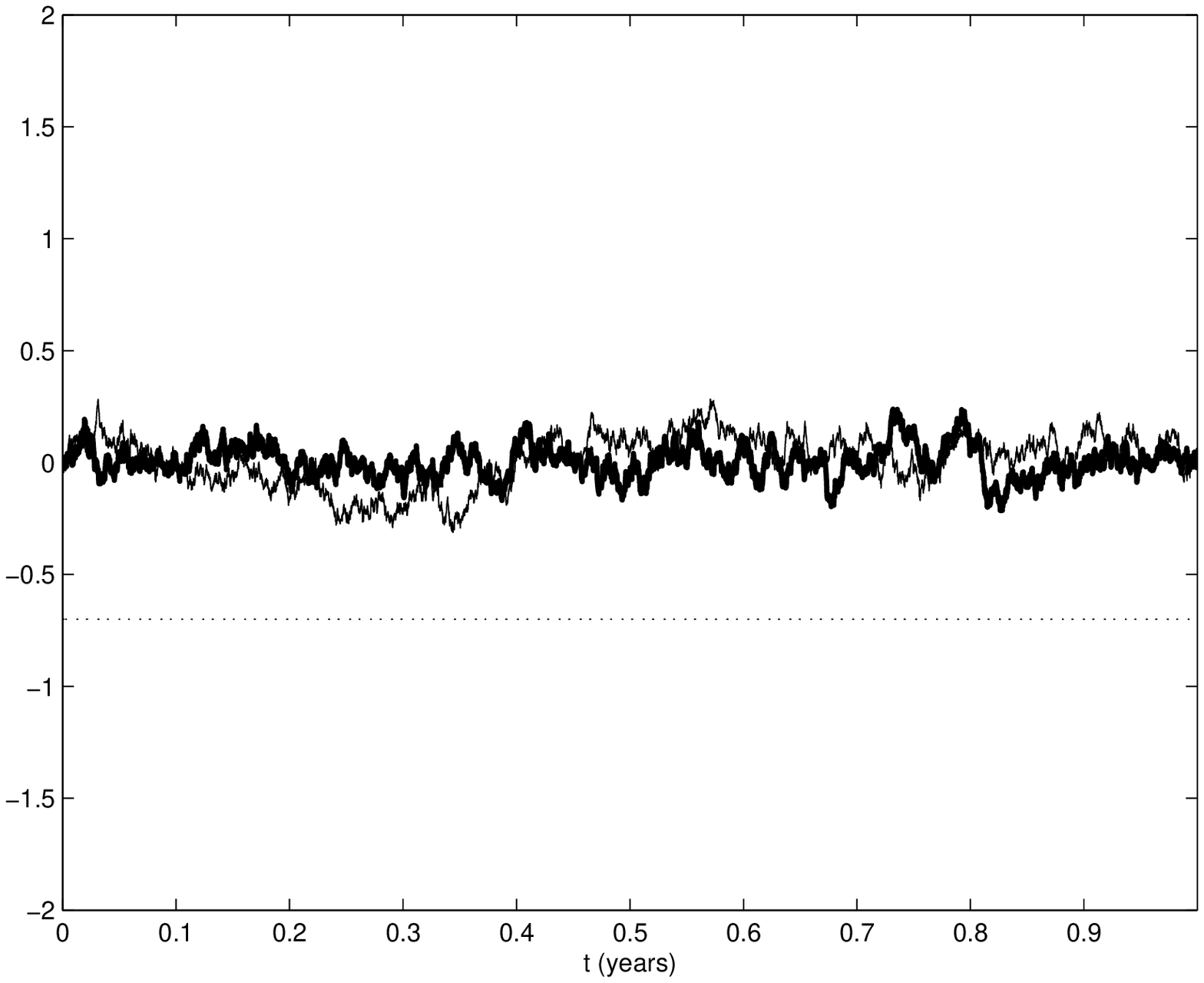}a)\includegraphics[height=6cm]{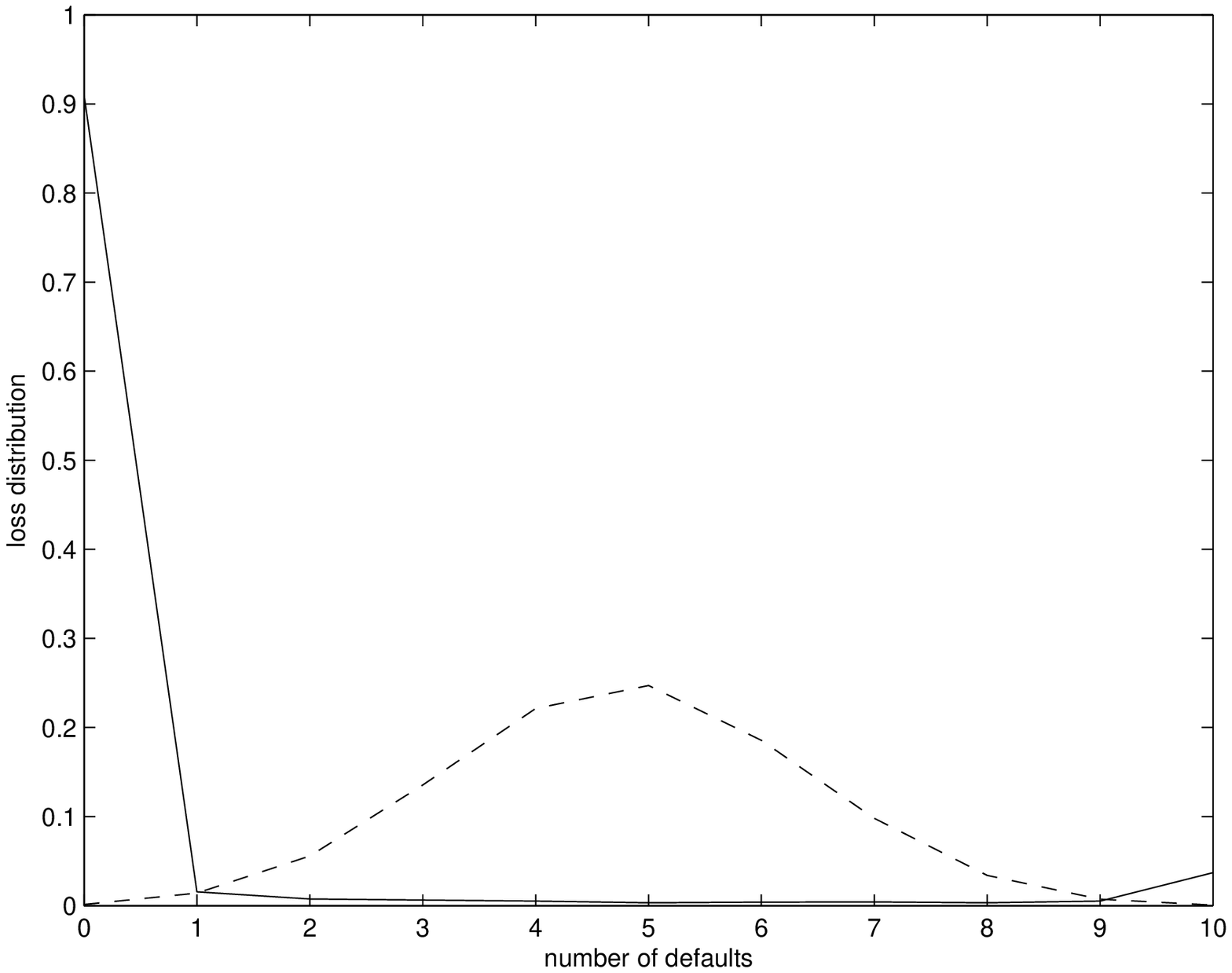}b)}
 \caption{\footnotesize 
 The solid line shows  one trajectory  of the  solution $Y_{t}^{1}$,  $ t \in [0,T]$, $T=1$,   of  system  (\ref{FBanks}), (\ref{FBanksCI}), (\ref{rho})     when 
  $N=10$,  $\alpha=100$. 
 The thick solid line   shows  one trajectory  of the  solution  ${\cal Y}_{t}$,  $ t \in [0,T]$, $T=1$,   of  the  corresponding mean field initial value problem   (\ref{FMF}), (\ref{FMFCI}).
  The dotted  line shows the default level $D=-0.7$. 
 b)  Loss distribution in  $[0,T]$,  $T=1$,   of system  (\ref{FBanks}), (\ref{FBanksCI}), (\ref{rho}) when  $N=10$,   $\alpha=100$  (solid line) and  loss distribution in  $[0,T]$,  $T=1$,  of   system  (\ref{indBanks}), (\ref{FBanksCI}),  (\ref{rho}) when $N=10$ (dashed line).}
\end{figure}

 \begin{figure}
 \centerline{\includegraphics[height=6cm]{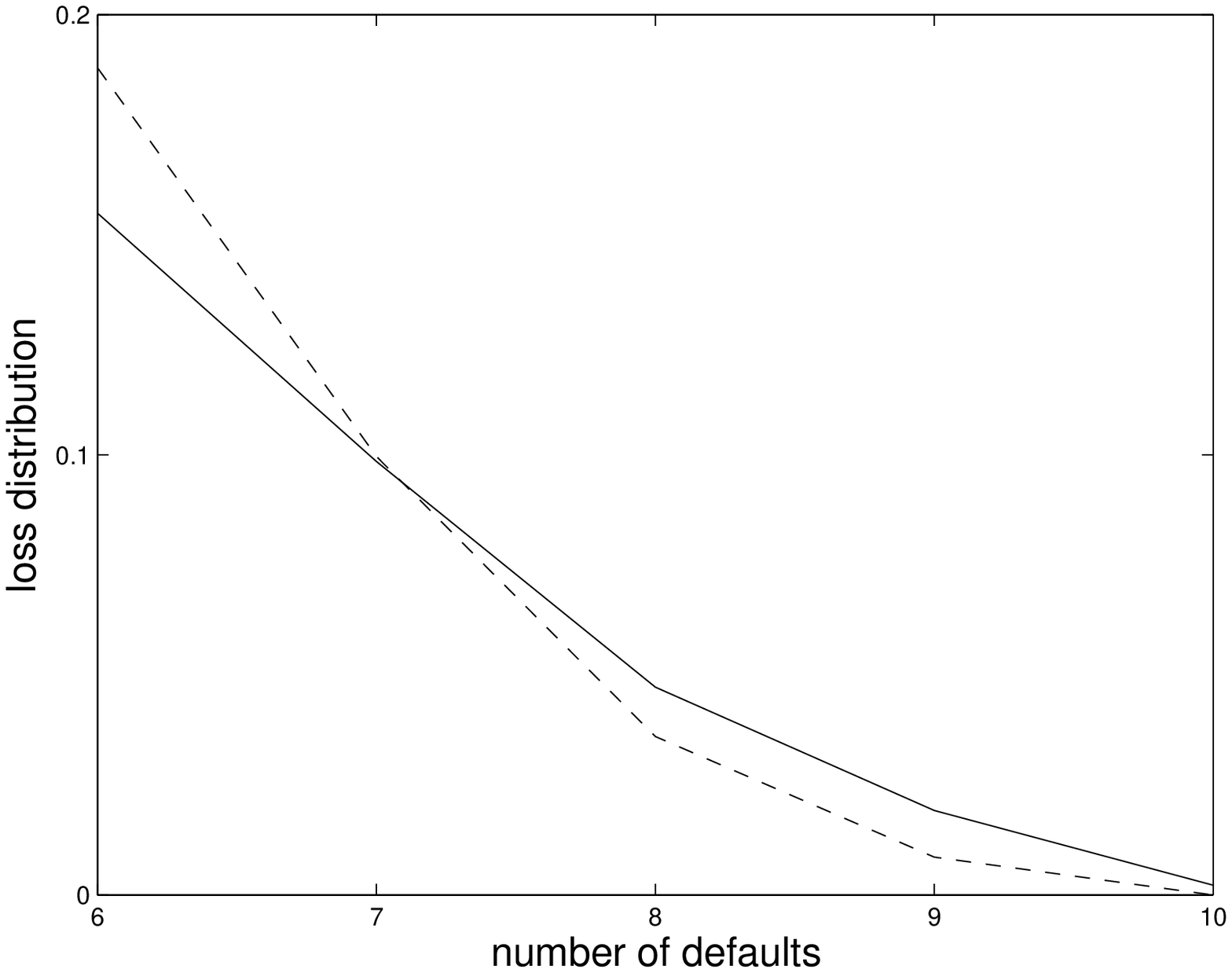}a)\includegraphics[height=6cm]{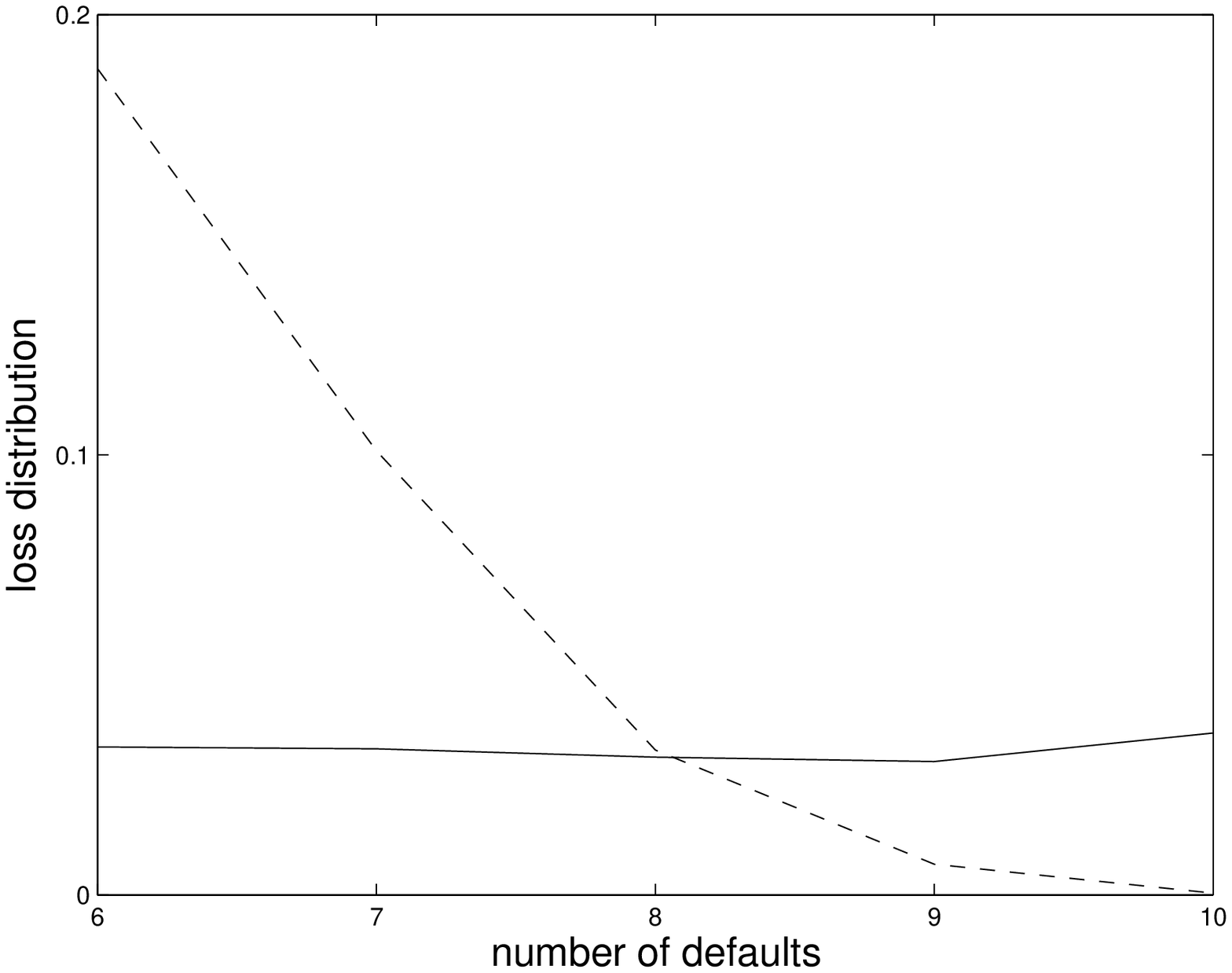}b)} \centerline{\includegraphics[height=6cm]{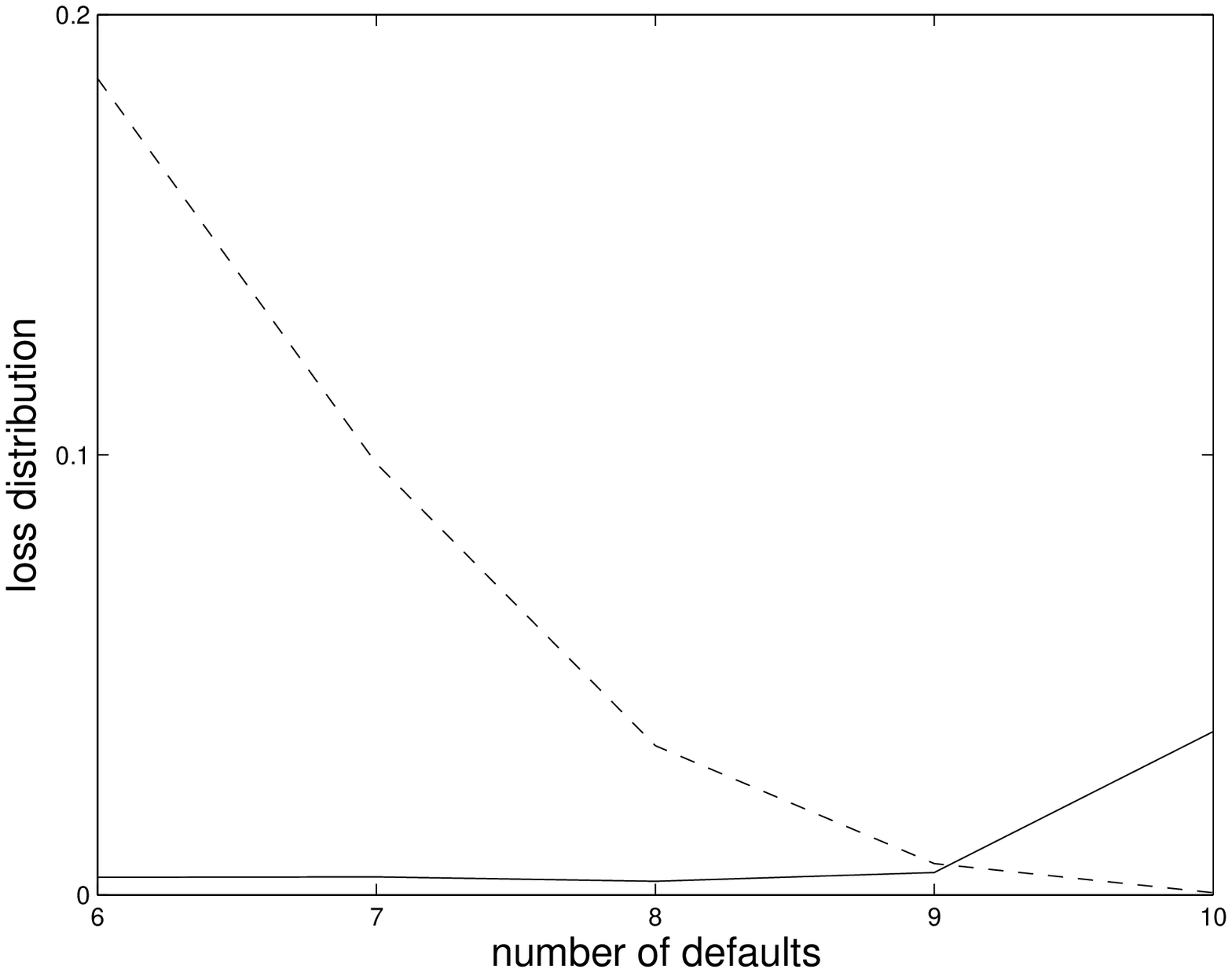}c)}
 \caption{\footnotesize 
 a) Zoom  of the  right-bottom corner of Figure 1b) ($\alpha=1$). 
 b) Zoom  of the  right-bottom corner of Figure 2b)  ($\alpha=10$). 
 c) Zoom  of the  right-bottom corner of Figure 3b)  ($\alpha=100$). 
 }
  \label{fig4}
\end{figure}

 %
 
%
\section{A banking system model with  two cooperation   mechanisms}\label{sec3}
%

%

Let us introduce the  banking system model  studied in this paper.
 The log-monetary reserves as  functions  of time  of the  $N$ banks present in the banking system  are modeled by  the    diffusion processes
 $X_{t}^{i}$, $t>0$,  $i=1,2,\ldots,N$.
   These diffusion processes are  defined implicitly  through an initial value problem for a  system of stochastic differential equations.    In the model   we  use  the cooperation   mechanism  present in   (\ref{FBanks}) to describe    the  inter-bank borrowing and lending activity, and     we introduce a new cooperation   mechanism  to describe  the borrowing and lending  activity between  banks and  monetary authority. 
 This last mechanism depends from the difference   as  function of time between   the empirical mean of the log-monetary reserves  of the banks    and a target  trajectory   $\xi_{t}=\xi(t)$,  $ t\ge0$, chosen by the  monetary authority that represents the log-monetary reserves   of the  ``ideal bank'' as a function of time. 
   
Given $\alpha\ge0$, $\gamma\le0$,  $\sigma>0$,  we begin  modeling   the dynamics of  the log-monetary reserves  of the banks  with    the following  system of  stochastic differential  equations:
\begin{eqnarray}
&& dX_{t}^{i}= \frac{\alpha}{N}  \sum_{j=1}^{N} \left( X_{t}^{j} -X_{t}^{i}\right) dt+ 
                          \frac{ \gamma }{N} \sum_{j=1}^{N} \left(X_{t}^{j} -\xi_{t}\right)  dt +d\xi_{t} + \sigma dW_{t}^{i},\nonumber\\
&& \hskip8truecm   t>0, \,  i=1,2,\ldots,N ,  \label{OurBanks}
\end{eqnarray}
with the  initial condition:
\begin{eqnarray}
&&X_{0}^{i} =\xi_{0}, \quad i=1,2,\ldots,N,                  \label{OurBanksCI}
\end{eqnarray}
where     $\xi_{t}=\xi(t)$,  $ t\ge0$,  is  a continuous piecewise differentiable function and    
 $\displaystyle d\xi_{t}=\frac{d\xi_{t}}{dt} \,dt=(\xi_{t})' \,dt$, $ t\ge0$.   The initial value problem  (\ref{OurBanks}),   (\ref{OurBanksCI}) is completed with the assumption (\ref{rho}).  
 Note that       with abuse       of notation    in (\ref{OurBanksCI})     $\xi_{0}$  is   a  random variable  concentrated in  the  point  $\xi_{t=0}$,  with
probability one. 
 We assume $\xi_{0} \ge 0$ and $\xi_{t}-D>0$, $t>0$, where $D$ is the default barrier.

 In (\ref{OurBanks})  the parameter $\alpha \ge 0 $  regulates   the first cooperation   mechanism expressed by the terms $\displaystyle  \frac{\alpha}{N} \sum_{j=1}^{N} \left( X_{t}^{j} -X_{t}^{i}\right) dt$, $t>0$, $i=1,2,\ldots,N, $  that  controls  the   inter-bank borrowing and lending activity (see  Section \ref{sec2}). 
 The term $d\xi_{t}$, $t>0$,     of  (\ref{OurBanks}) is a part of  the intervention of the monetary authority in the banking system dynamics and   is responsible for the fact  that the drift terms  $\displaystyle  \frac{\alpha}{N} \sum_{j=1}^{N} \left( X_{t}^{j} -X_{t}^{i}\right) dt$, $t>0$, $i=1,2,\ldots,N, $ stabilize  the trajectories   of $X_{t}^{i}$, $t>0$,  $i=1,2,\ldots,N, $ around  $\hat\xi_{t}=\xi_{t}$, $t>0$, instead than around $\hat\xi_{t}=\xi_{0}$, $t>0$, as done in (\ref{FBanks}), (\ref{FBanksCI}) where  we have    $\xi_{0}=0$ (see Section \ref{sec2}).
In  (\ref{OurBanks}) the parameter $\gamma \le 0  $  regulates the second cooperation  mechanism   expressed by the term  $\displaystyle   \frac{ \gamma }{N} \sum_{j=1}^{N} \left(X_{t}^{j} -\xi_{t}\right)  dt$, $t>0$, added to the $i$-th equation, $i=1,2,\ldots,N$.
 That is $\gamma$     controls the strength of mean reversion   of   the empirical mean of the  log-monetary reserves  of the banks   $\displaystyle \frac{1}{N} \sum_{j=1}^{N} X_{t}^{j}$  to the target  trajectory  of the  log-monetary reserves of the  ``ideal bank''   $\xi_{t}$, $t>0$.  
Equations (\ref{OurBanks})   through the term  proportional to  $\gamma$ impose that the  mean of the log-monetary reserves  of the banks of the system (i.e. $\displaystyle \frac{1}{N} \sum_{j=1}^{N} X_{t}^{j}$, $t>0$) reverts to the  log-monetary reserves of the  ``ideal bank'' (i.e. $\xi_{t}$,  $ t>0$).  In the model  (\ref{OurBanks}),   (\ref{OurBanksCI}), (\ref{rho}) the  bank population  is homogeneous
and  when $N$ goes to infinity  each bank behaves like  the mean bank. 
That is    the target trajectory  $\xi_{t}$, $t>0$,  represents  the ``ideal'' log-monetary reserves  of each  bank and  $N\xi_{t}$, $t>0$,  represents  the ``ideal'' target  trajectory  of the   log-monetary reserves of the banking system. 
The parameter $\gamma$  controls  the rate of  the borrowing and  lending activity  between  banks and   monetary authority.
That is the  second  cooperation  mechanism  of the model is regulated by  $\gamma$ and  stabilizes  the empirical mean of the diffusion processes   $X_{t}^{i}$, $t>0$,  $i=1,2,\ldots,N, $ along  the target trajectory $\xi_{t}$, $t>0$. 
 In fact    the term   $ \displaystyle\frac{ \gamma }{N} \sum_{j=1}^{N} \left(X_{t}^{j} -\xi_{t}\right)  dt$, $t>0$,  of (\ref{OurBanks})  can be rewritten as  $ \displaystyle \gamma \left( \frac{ 1 }{N} \sum_{j=1}^{N} X_{t}^{j} -\xi_{t}\right)  dt$, $t>0$.

Let us derive  the mean field approximation of the  banking system  model (\ref{OurBanks}),   (\ref{OurBanksCI}), (\ref{rho}).
%
%
First of all let us  rewrite   the  system of equations  (\ref{OurBanks})  as follows:
\begin{eqnarray}
&& dX_{t}^{i}= \alpha\left[ \left(  \frac{1}{N}  \sum_{j=1}^{N}  X_{t}^{j} \right) -X_{t}^{i} \right] dt+ 
                         \gamma \left( \frac{1}{N} \sum_{j=1}^{N} X_{t}^{j} -\xi_{t}\right)  dt +d\xi_{t} + \sigma dW_{t}^{i},\nonumber\\
&& \hskip8truecm   t>0, \,  i=1,2,\ldots,N .  \label{step1}
\end{eqnarray}
Summing  the  equations (\ref{step1}) for $i$  going from $1$ to $N$, and dividing the resulting equation   by N we have:
\begin{eqnarray}
 d\left(  \frac{1}{N}  \sum_{i=1}^{N}  X_{t}^{i} \right) = 
                         \gamma \left( \frac{1}{N} \sum_{j=1}^{N} X_{t}^{j} -\xi_{t}\right)  dt +d\xi_{t} + 
                         d\left(\frac{\sigma}{N} \sum_{i=1}^{N}W_{t}^{i}\right), \quad  t>0. \label{step2}
\end{eqnarray}
Let  
\begin{equation}\label{mean}
 {\bar X}_{t}= \frac{1}{N} \sum_{j=1}^{N} X_{t}^{j} , \quad  t>0, 
\end{equation}
be  the empirical mean of the processes $X_{t}^{j},$ $  t>0$, $j=1,2,\ldots,N$.  Equation (\ref{step2}) can be rewritten as follows:
\begin{eqnarray}
&& d{\bar X}_{t} = 
                         \gamma \left({\bar X}_{t}  -\xi_{t}\right)  dt +d\xi_{t} + 
                         d\left(\frac{\sigma}{N} \sum_{i=1}^{N}W_{t}^{i}\right),  \quad  t>0. \label{step2bis}
\end{eqnarray}
The assumption   (\ref{rho}) and   the strong law of large numbers imply  that:
\begin{eqnarray}
 \frac{1}{N} \sum_{i=1}^{N}W_{t}^{i} \longrightarrow 0, \quad t>0, \quad  \mbox{almost surely},  \, \mbox{when}  \, \, N   \longrightarrow +\infty. \label{step3}
\end{eqnarray}
Let  $ \mediaX_{t}$, $t>0$,  be  the limit  as  $N$  goes to infinity   of the empirical mean $ {\bar X}_{t}$, $t>0$, and    ${\cal X}_{t}$, $t>0,$ be  the  process representing  the  log-monetary reserves of the ``mean-bank'' of the banking  system model (\ref{OurBanks}),   (\ref{OurBanksCI}), (\ref{rho}) in the limit  $N$  goes to infinity.
The mean field approximation  of the  dynamical system     (\ref{OurBanks}),   (\ref{OurBanksCI}), (\ref{rho})   can be deduced   immediately  from (\ref{step1}),  (\ref{step2bis}). In fact   when  $N$  goes to infinity from (\ref{step1}),  (\ref{step2bis})   we have:
\begin{eqnarray}
&&  d{\cal X}_{t}= \alpha \left(   \mediaX_{t} - {\cal X}_{t} \right) dt+ \gamma \left(   \mediaX_{t} - \xi_{t} \right) dt +d\xi_{t} + \sigma dW_{t},\quad t>0, \label{OurMF1}\\
&&       d \mediaX_{t}   =\gamma \left(   \mediaX_{t} - \xi_{t} \right) dt +d\xi_{t} , \quad t>0    ,      \label{OurMF2}\\
&&       {\cal X}_{0}   =\xi_{0},    \quad     \mediaX_{0}   =\xi_{0}.  \label{OurMF3}
\end{eqnarray}
The initial value problem (\ref{OurMF1}),  (\ref{OurMF2}),  (\ref{OurMF3}) is the mean field approximation  of the  dynamical system     (\ref{OurBanks}),   (\ref{OurBanksCI}), (\ref{rho}).
%
%
Note that equations (\ref{OurMF2}) and (\ref{OurMF3}) imply that   $\mediaX_{t}=\xi_{t}$, $t>0$,  therefore    in (\ref{OurMF1})  $\gamma$ is multiplied by zero.
   This means that  the choice of    $\gamma$  does not influence   the trajectories of  (\ref{OurMF1}).  That   is  the second cooperation   mechanism    introduced in  (\ref{OurBanks})  does not contribute to   the mean field approximation  (\ref{OurMF1}),  (\ref{OurMF2}),  (\ref{OurMF3}) of  (\ref{OurBanks}),   (\ref{OurBanksCI}), (\ref{rho}).      This is  undesirable since  we  want     to use  the mean field approximation   (\ref{OurMF1}),  (\ref{OurMF2}),  (\ref{OurMF3})  to determine the parameters  $\alpha$, $\gamma$   that must be used 
  to  govern  the  banking system model  (\ref{OurBanks}),   (\ref{OurBanksCI}), (\ref{rho}).   
 To  avoid this inconvenience   let us  modify   the  banking system model (\ref{OurBanks}),   (\ref{OurBanksCI}), (\ref{rho}).
 Starting from  the target trajectory  $\xi_{t}$, $t\ge 0$, we introduce  two  auxiliary target trajectories  $\xi_{t}^{-}$, $\xi_{t}^{+}$, $t \ge 0$,  such that $\xi_{t}^{-}\ne\xi_{t}^{+}$, $t \ge 0$,  obtained from $\xi_{t}$, $t \ge 0$, with  a ``slight'' perturbation.  For example, given $\epsilon >0$,  we  choose    $\xi_{t}^{-}$, $\xi_{t}^{+}$, $t \ge 0$, as follows:
\begin{eqnarray} \label{xi12}
\xi_{t}^{-}=\xi_{t}-\epsilon, \quad t\ge0, \qquad \qquad \qquad \xi_{t}^{+}=\xi_{t}+\epsilon, \quad t \ge 0.
\end{eqnarray}
We replace (\ref{OurBanks}),   (\ref{OurBanksCI})  with  the  system of stochastic differential  equations:
\begin{eqnarray}
&& dX_{t}^{i}= \frac{\alpha}{N}  \sum_{j=1}^{N} \left( X_{t}^{j} -X_{t}^{i}\right) dt+ 
                          \gamma \left( \frac{1}{N} \sum_{j=1}^{N} X_{t}^{j} -\xi_{t}^{-}\right)  dt +d\xi_{t}^{+} + \sigma dW_{t}^{i},\nonumber\\
&& \hskip8truecm   t>0, \,  i=1,2,\ldots,N ,  \label{OurBanks12}
\end{eqnarray}
 with the initial condition:
\begin{eqnarray}
&&X_{0}^{i} =\xi_{0}^{+}, \quad i=1,2,\ldots,N,                  \label{OurBanksCI12}
\end{eqnarray}
where, with abuse of notation,  $\xi_{0}^{+}$     denotes a random variable   concentrated in the point   $\xi_{t=0}^{+}$  with probability one.  The initial value problem  (\ref{OurBanks12}),   (\ref{OurBanksCI12}) is completed with the assumption (\ref{rho}). 

The system  of equations (\ref{OurBanks12}),   (\ref{OurBanksCI12}), (\ref{rho})  is the  banking system model with  two cooperation   mechanisms   studied     in this paper.  
The  two cooperation  mechanisms of  model (\ref{OurBanks12}),   (\ref{OurBanksCI12}), (\ref{rho})  are   the same  of   those   of   model  (\ref{OurBanks}),   (\ref{OurBanksCI}), (\ref{rho}). 
However  in   model  (\ref{OurBanks12}),   (\ref{OurBanksCI12}), (\ref{rho}) the  inter-bank borrowing and lending mechanism, governed by $\alpha$,  stabilizes the system around the trajectory $\xi_{t}^{+}$, $t \ge 0$,   and    the banks-monetary authority  borrowing and lending mechanism, governed by $\gamma$,  stabilizes the system around the trajectory $\xi_{t}^{-}$, $t \ge 0$, and   we have $\xi_{t}^{+}\ne \xi_{t}^{-}$, $t \ge 0$.

Proceeding  as done   previously  in the study of   model  (\ref{OurBanks}),   (\ref{OurBanksCI}), (\ref{rho})    it is easy to see  that the mean field 
approximation of the dynamical system    (\ref{OurBanks12}),   (\ref{OurBanksCI12}), (\ref{rho})  is:
\begin{eqnarray}
&&  d{\cal X}_{t}= \alpha \left(   \mediaX_{t} - {\cal X}_{t} \right) dt+ \gamma \left(   \mediaX_{t} - \xi_{t}^{-} \right) dt +d\xi_{t}^{+} + \sigma dW_{t},\quad t>0, \label{12OurMF1}\\
&&       d \mediaX_{t}   =\gamma \left(   \mediaX_{t} - \xi_{t}^{-} \right) dt +d\xi_{t}^{+} , \quad t>0    ,      \label{12OurMF2}\\
&&       {\cal X}_{0}   =\xi_{0}^{+},    \quad     \mediaX_{0}   =\xi_{0}^{+},  \label{12OurMF3}
\end{eqnarray}
where ${\cal X}_{t}$, $t>0,$ is a stochastic process that represents the time evolution of the log-monetary reserves of the ``mean-bank'' of  model (\ref{OurBanks12}),   (\ref{OurBanksCI12}), (\ref{rho}) and $ \mediaX_{t}$, $t>0,$ is an auxiliary variable.

For simplicity   we use the same variables   
$X_{t}^{i}$, $ i=1,2,\ldots,N, $   $ {\bar X}_{t}$,  ${\cal X}_{t}$, $ \mediaX_{t}$, $t>0$,   to denote  the dependent variables of   the  models   (\ref{OurBanks}),   (\ref{OurBanksCI}), (\ref{rho}) and (\ref{OurBanks12}),   (\ref{OurBanksCI12}), (\ref{rho}) and of  the  corresponding mean field equations  (\ref{OurMF1}), (\ref{OurMF2}), (\ref{OurMF3}) and  (\ref{12OurMF1}), (\ref{12OurMF2}), (\ref{12OurMF3}).

Let us point out  that the models   (\ref{OurBanks}),   (\ref{OurBanksCI}), (\ref{rho}) and   (\ref{OurBanks12}),   (\ref{OurBanksCI12}), (\ref{rho}) can  both    be governed using (\ref{12OurMF1}), (\ref{12OurMF2}), (\ref{12OurMF3}).  
In fact instead of  (\ref{OurBanks12}),   (\ref{OurBanksCI12}), (\ref{rho}) we can  consider  (\ref{OurBanks}),   (\ref{OurBanksCI}), (\ref{rho}) as banking system model   and we can   use the mean field equations (\ref{12OurMF1}), (\ref{12OurMF2}), (\ref{12OurMF3})   as shown in Section \ref{sec4} to govern 
(\ref{OurBanks}),   (\ref{OurBanksCI}), (\ref{rho}). 
We prefer to choose     (\ref{OurBanks12}),   (\ref{OurBanksCI12}), (\ref{rho})   as banking system   model   since  with this choice the  banking system model    is governed  through its  mean field equations. 
Model    (\ref{OurBanks}),   (\ref{OurBanksCI}), (\ref{rho})  can be  governed with the method presented in Section \ref{sec4}
through the auxiliary  dynamical system   (\ref{12OurMF1}), (\ref{12OurMF2}), (\ref{12OurMF3})  but  cannot be governed  using   its mean  field approximation   (\ref{OurMF1}), (\ref{OurMF2}), (\ref{OurMF3}) since   in  (\ref{OurMF1})  $\gamma$ is multiplied by zero.

%
For later purposes  in the   banking system model       (\ref{OurBanks12}),   (\ref{OurBanksCI12}), (\ref{rho})  and  in its   mean field approximation   (\ref{12OurMF1}), (\ref{12OurMF2}), (\ref{12OurMF3})  from now on   we allow the  parameters  $\alpha$ and $\gamma$  to  be functions of time.   That is   in   (\ref{OurBanks12}),   (\ref{OurBanksCI12}), (\ref{rho})  and in  (\ref{12OurMF1}), (\ref{12OurMF2}), (\ref{12OurMF3}) we  assume:
\begin{eqnarray} \label{alphagammatime} 
&& 
\alpha=\alpha_{t} \ge 0,  \quad  t\ge 0, \qquad  \gamma=\gamma_{t} \le 0,  \quad  t\ge 0.
\end{eqnarray}
It is easy to see that the   functioning of the  two cooperation  mechanisms present  in (\ref{OurBanks12}),   (\ref{OurBanksCI12}), (\ref{rho})   
remains unchanged  when  we consider    $\alpha=\alpha_{t}\ge 0,$ $t\ge 0$,  $\gamma=\gamma_{t}\le 0,$ $t\ge 0$,  to be   functions of time. 
 
Instead of choosing $\xi_{t}$, $t \ge 0$, and $\epsilon$  the monetary authority can    choose  an interval  of the real axis    as function of time, that is  the interval $\left[\tilde\xi_{t}^{-} , \tilde\xi_{t}^{+}\right]$, $t \ge 0$,    defined by  two  target trajectories  $\tilde\xi_{t}^{-}$, $\tilde\xi_{t}^{+}$, $t \ge 0$, such that  $\tilde\xi_{t}^{-} < \tilde\xi_{t}^{+}$, $t \ge 0$. In  this case    $\xi_{t}$, $t \ge 0$,  can  be defined as the arithmetic  mean of $\tilde\xi_{t}^{-}$, $\tilde\xi_{t}^{+}$, $t \ge 0$.



 Figures  5a)-7a)  show one trajectory of  $X_{t}^{1}$,  $ t \in [0,T]$, $T=1$, solution  of  model   (\ref{OurBanks12}),   (\ref{OurBanksCI12}), (\ref{rho}),  and one trajectory  of  ${\cal X}_{t}$,  $ t \in [0,T]$, $T=1$,  solution of the  initial value problem   (\ref{12OurMF1}), (\ref{12OurMF2}), (\ref{12OurMF3}), 
 when 
 $\xi(t)=\displaystyle 0.5 \sin (2 \pi t)$, $ t \in [0,T]$, $T=1$,  $\epsilon=0.05$,  and  the constants  $\alpha$, $\gamma$   are  chosen as follows: 
  $\alpha=100,$  $\gamma=-10$ (see Figure 5a)),  $\alpha=50$, $\gamma=-50$  (see Figure 6a)) and $\alpha=10$, $\gamma=-100$ (see Figure 7a)).  As    done in  Section \ref{sec2} we choose $N=10$,  $D=-0.7$, $\sigma=1$ and we use the explicit Euler method   with time step $\Delta t=10^{-4} $ to integrate numerically the stochastic differential models (\ref{OurBanks12}),   (\ref{OurBanksCI12}), (\ref{rho})  and (\ref{12OurMF1}), (\ref{12OurMF2}), (\ref{12OurMF3}),
  for $ t \in [0,T]$, $T=1$.
 In  Figures  5a)-7a)  the  dotted line shows  the default level $D=-0.7$,  the dashed   line  shows  the target  trajectory  $\xi(t)=\displaystyle 0.5 \sin (2 \pi t)$, $ t \in [0,T]$, $T=1$.
 In Figures 5a)-7a)  we  show  only  one trajectory of $X_{t}^{1}$,  $ t \in [0,T]$, $T=1$,  solution of model  (\ref{OurBanks12}),   (\ref{OurBanksCI12}), (\ref{rho}) since the bank population  described by  (\ref{OurBanks12}),   (\ref{OurBanksCI12}), (\ref{rho})  is homogeneous and the processes $X_{t}^{i}$,  $ t \in [0,T]$, $T=1$, $i=1,2,\ldots,N$, behave all  in the same way.  
Let us  study the loss distribution  in  the time interval   $[0,T]$,  $T=1$,  of the diffusion processes solution of    (\ref{OurBanks12}), (\ref{OurBanksCI12}), (\ref{rho}), that is let us consider the probability distribution  of the random variable: number of bank defaults   in the  time interval  $[0,T]$,  $T=1$.  The loss distribution in   $[0,T]$,  $T=1$,  is evaluated   using  statistical simulation on a set of   $10^{4}$ numerically generated trajectories of  (\ref{OurBanks12}), (\ref{OurBanksCI12}), (\ref{rho}). 
 In Figures  5b)-7b) the dashed line represents the loss distribution of the independent  diffusion processes  solution  of (\ref{indBanks}), (\ref{FBanksCI}),  (\ref{rho}) when $N=10$,  that is  of the solution of system   (\ref{OurBanks12}), (\ref{OurBanksCI12}), (\ref{rho})  when  $\alpha=\gamma=0$,   $N=10$, $\xi(t)=0$, $ t \in [0,T]$, $T=1$, $\epsilon=0$, while the solid  line represents  the loss distribution of   the  diffusion processes solution of system  (\ref{OurBanks12}), (\ref{OurBanksCI12}), (\ref{rho}) 
 when $N=10$,  $\xi(t)=\displaystyle 0.5 \sin (2 \pi t)$, $ t \in [0,T]$, $T=1$, $\epsilon=0.05$,  for the following values of the parameters  $\alpha$ and $\gamma$:  $\alpha=100,$  $\gamma=-10$ (Figure 5b)),  $\alpha=50,$  $\gamma=-50$ (Figure 6b)) and $\alpha=10,$  $\gamma=-100$ (Figure 7b)). 

Figures 5-7 show how the banking system model   (\ref{OurBanks12}),   (\ref{OurBanksCI12}), (\ref{rho})   is stabilized  by the two cooperation   mechanisms discussed previously.  In particular  Figures 5-7 show the different effects of these  two mechanisms. 
When $\alpha$ and $\gamma$   are  constants  and $\alpha$  dominates over  $|\gamma|$, that is  when $\alpha>>|\gamma|$, (see Figure 5) we are in a situation similar  to the situation discussed  in Section \ref{sec2}    for the Fouque and Sun model  (\ref{FBanks}), (\ref{FBanksCI}),  (\ref{rho}) and presented  in Figures 1-3.  In this case the effect of the borrowing and lending activity among banks  is dominant. This  mechanism  when $\alpha$ increases
 improves the stability of the individual bank    at the expense of  an  increased    probability of  extreme  systemic risk. 
 Moreover  Figure 5  shows the ``swarming effect'' along the trajectory $\xi_{t}^{+}$,  $ t \in [0,T]$, $T=1$,   induced  by the term $d\xi_{t}^{+}$  in (\ref{OurBanks12}).
When   $\alpha$ and $\gamma$   are  constants and $|\gamma|$  dominates over $\alpha$, that is  when $\alpha<<|\gamma|$ (see Figure 7)   the  effect of the borrowing and lending  activity between  banks and   monetary authority is dominant.  This mechanism  stabilizes the empirical mean of the log-monetary reserves of the  banks  along the trajectory $\xi_{t}^{-}$,  $ t \in [0,T]$, $T=1$,    without changing  substantially   the probability  of  systemic risk. This stabilization effect increases when $|\gamma|$   increases. 
Note that the  stabilization of  the empirical mean of the log-monetary reserves   of the banks  around a ``safe'' target trajectory
is useful to  protect  the banking system from systemic failures.
In fact  empirical mean of the     log-monetary reserves  of the banks   stabilized around a ``safe'' target  trajectory means that if there is a failed  bank (i.e. a bank  with  log-monetary reserves   below  the  ``safe'' target  trajectory and in fact below the default level)
there must be a bank   or a  group of banks that are ``prosperous'' (i.e. banks with  log-monetary reserves  above the ``safe''  target trajectory).  In this situation   it is   possible for a   prosperous bank or for  a  consortium   of  prosperous banks to rescue  the failed  one or the failed  ones. In the intermediate  cases, that is  when $\alpha$ is of the same order of magnitude of $|\gamma|$  (see Figure 6)   the effects of the two mechanisms are  balanced. 

It is easy to  understand   that  given the default  level $D$   raising     the target trajectory $\xi_{t}$, $t>0$,  
(i.e. increasing  $\xi_{t}-D>0$, $t>0$)  decreases the  probability  of    systemic risk in a bounded time interval. 
However raising  the target trajectory  has the consequence  of    raising     the     log-monetary reserves of the banks.    This fact discourages economic activity.  Note that the loss of economic activity induced by the raise of  $\xi_{t}$, $t>0$, is an undesired effect  not considered in our model.
Similarly decreasing the target trajectory $\xi_{t}$, $t>0$,  (i.e. decreasing  $\xi_{t}-D>0$, $t>0$) increases the  probability  of    systemic risk in a bounded time interval and  has the consequence  of  decreasing  the  log-monetary reserves of the banks. 
We  conclude that the monetary authority must pursue  the goal of  avoiding systemic failures   keeping   the log-monetary reserves of the banks as small as possible.
  


%
 \begin{figure}
 \centerline{\includegraphics[height=6cm]{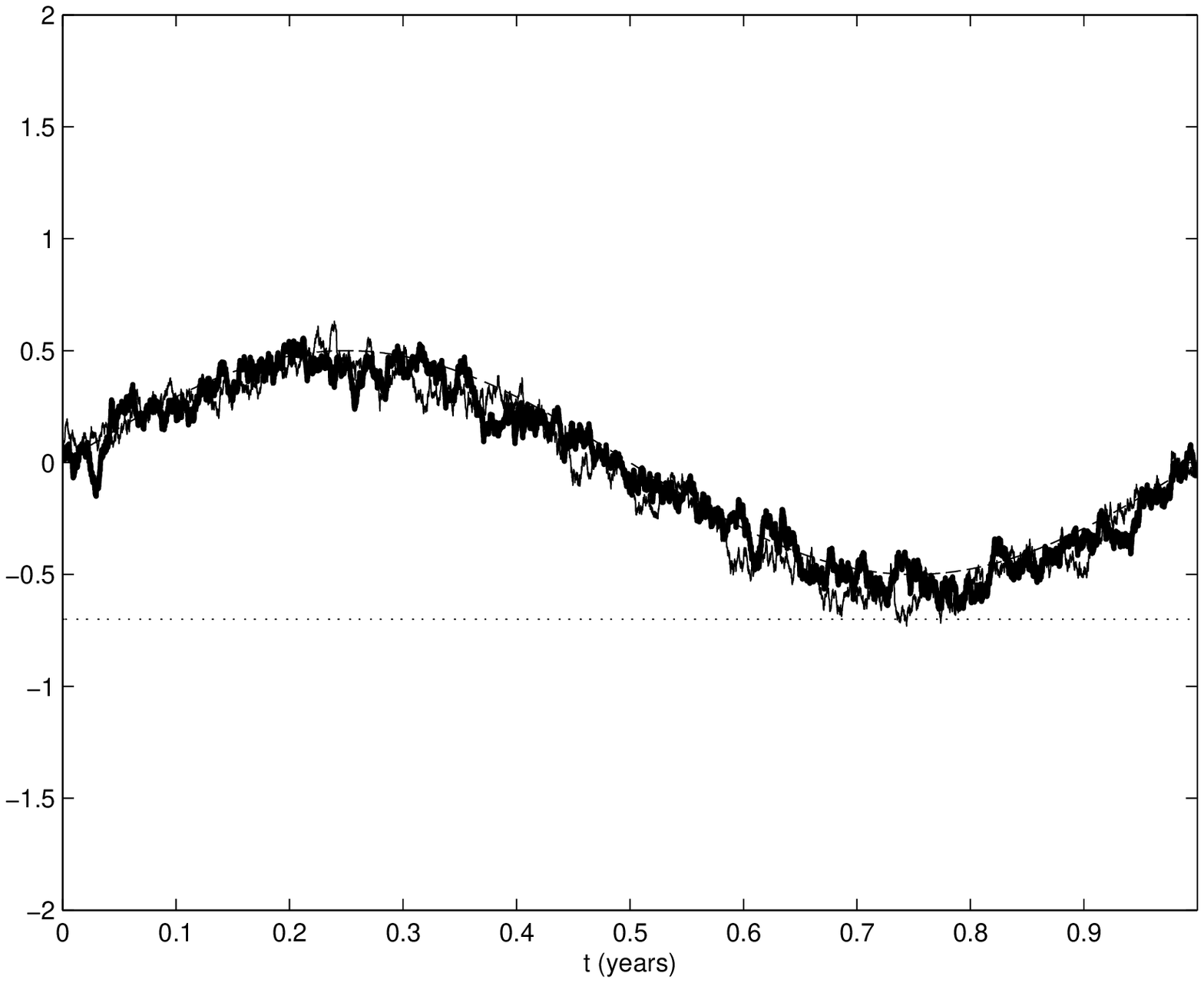}a)\includegraphics[height=6cm]{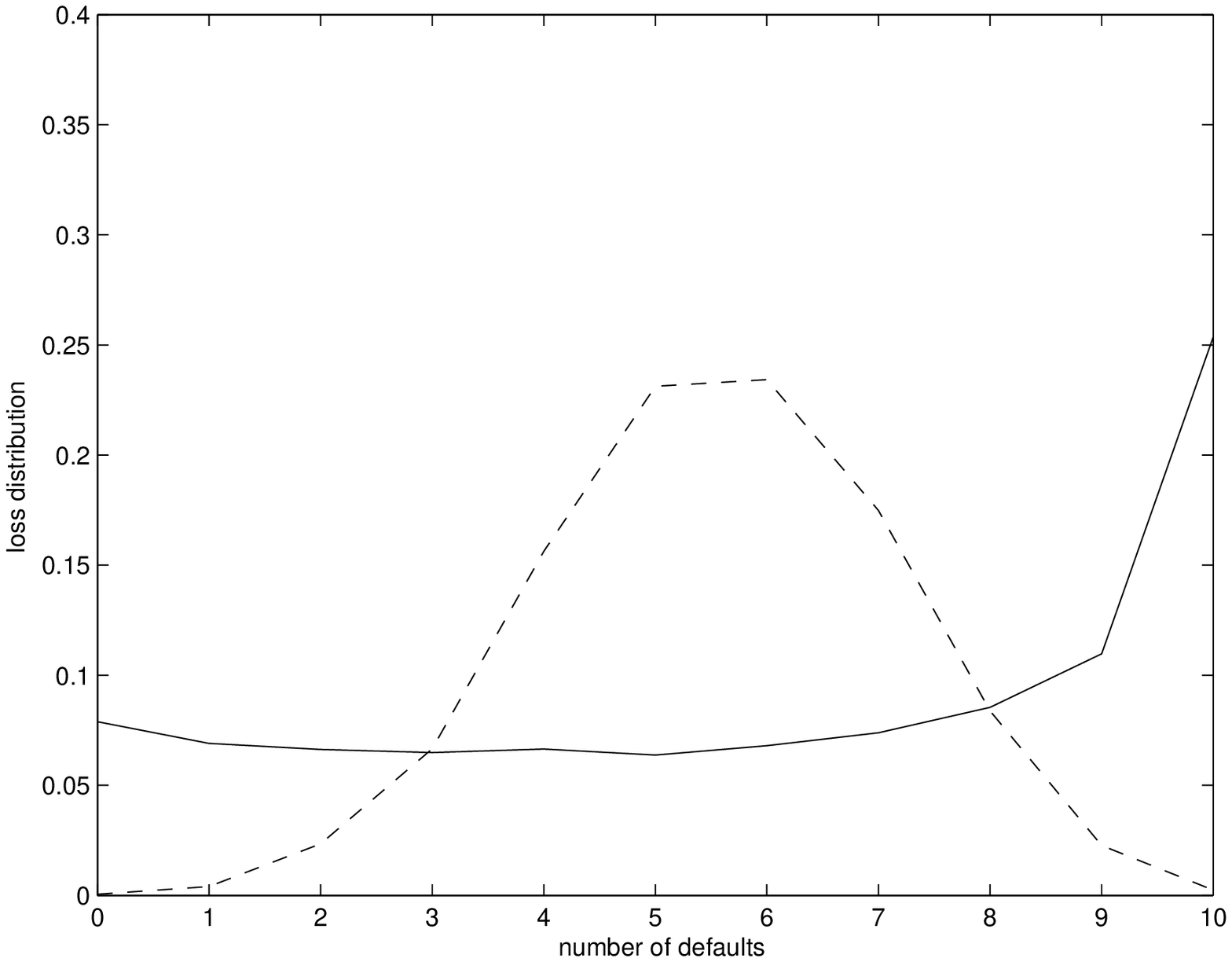}b)}
  \caption{\footnotesize 
 a) The solid line shows  one trajectory   of $X_{t}^{1}$,  $ t \in [0,T]$, $T=1$,   solution  of  system (\ref{OurBanks12}), (\ref{OurBanksCI12}), (\ref{rho}) when  $N=10$,   $\epsilon=0.05$, $\alpha=100,$  $\gamma=-10$.  
 The dashed line shows  the target  trajectory  $\xi(t)=\displaystyle 0.5 \sin (2 \pi t)$, $ t \in [0,T]$, $T=1$.
 The thick solid line   shows one trajectory  of the solution ${\cal X}_{t}$,  $ t \in [0,T]$, $T=1$, of the corresponding   mean field initial value problem  (\ref{12OurMF1}), (\ref{12OurMF2}), (\ref{12OurMF3}). 
 The dotted  line shows the default level $D=-0.7$. 
 b)  Loss distribution in  $[0,T]$,  $T=1$,  of system  (\ref{OurBanks12}), (\ref{OurBanksCI12}), (\ref{rho})  when $N=10$, $\epsilon=0.05$,  $\alpha=100$, $\gamma=-10$ (solid line) and  loss distribution in  $[0,T]$,  $T=1$,  of   system  (\ref{OurBanks12}), (\ref{OurBanksCI12}), (\ref{rho})  when $N=10$,  $\epsilon=0$,  $\alpha=\gamma=0$ (dashed  line).}
 \label{fig5}
 %
  \centerline{\includegraphics[height=6cm]{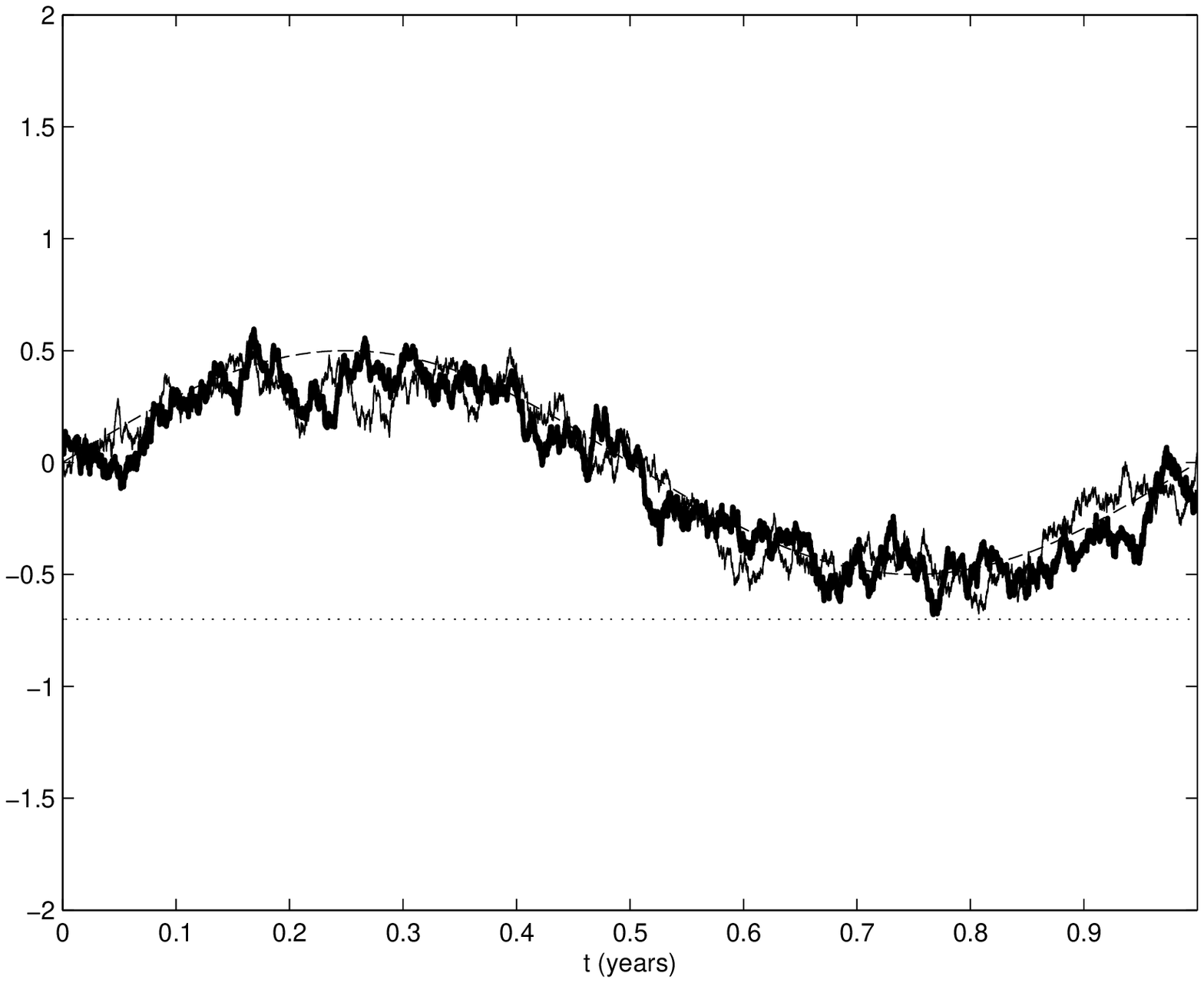}a)\includegraphics[height=6cm]{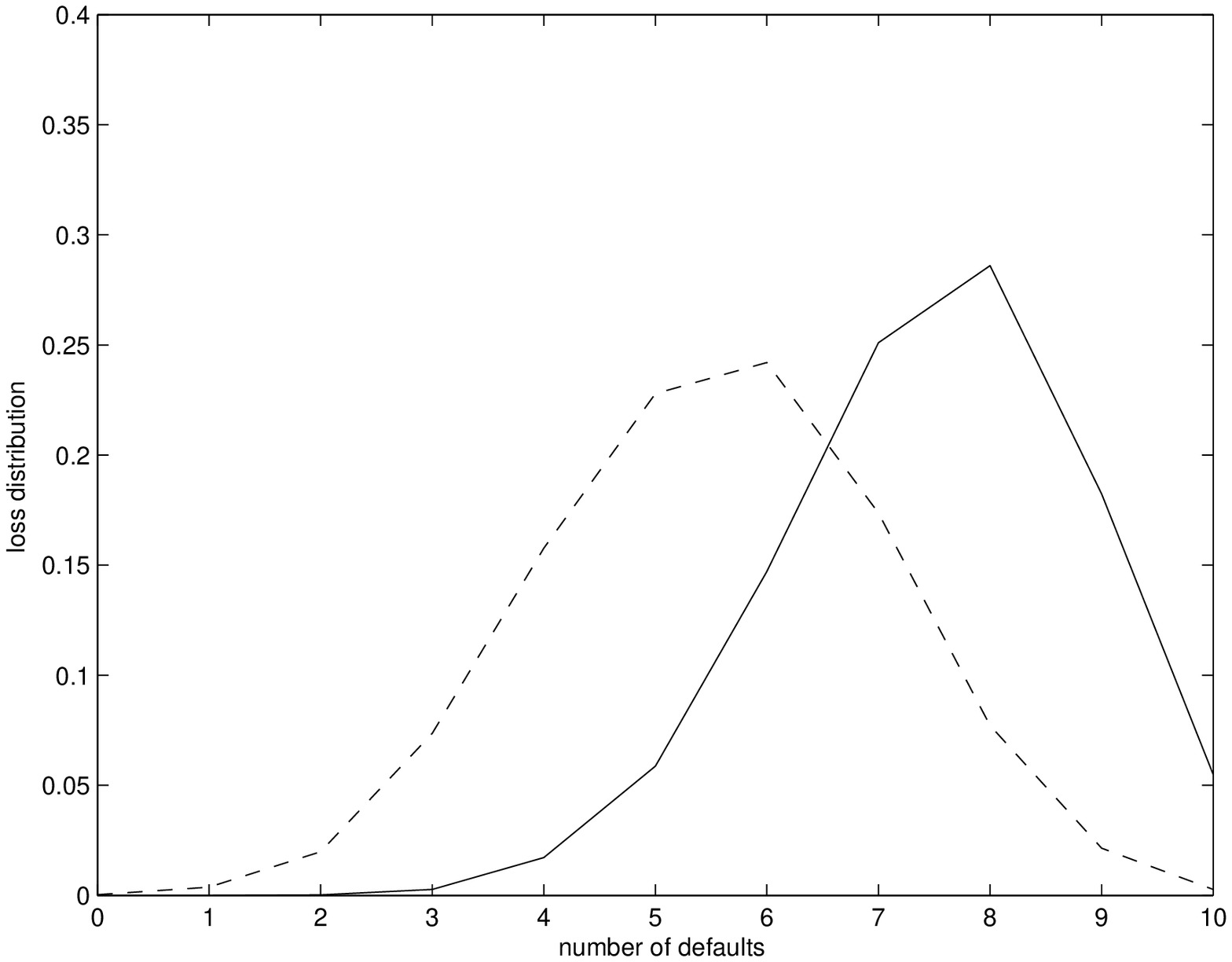}b)}
  \caption{\footnotesize 
  a) The solid line shows  one trajectory  of the  solution $X_{t}^{1}$,  $ t \in [0,T]$, $T=1$,  of  system (\ref{OurBanks12}), (\ref{OurBanksCI12}), (\ref{rho}) when $N=10$,    $\epsilon=0.05$, $\alpha=50,$  $\gamma=-50$.  
  The dashed line shows  the target  trajectory  $\xi(t)=\displaystyle 0.5 \sin (2 \pi t)$, $ t \in [0,T]$, $T=1$.
 The thick solid line   shows  one  trajectory of  ${\cal X}_{t}$,  $ t \in [0,T]$, $T=1$,  solution of the corresponding   mean field initial value problem  (\ref{12OurMF1}), (\ref{12OurMF2}), (\ref{12OurMF3}). 
 The dotted  line shows the default level $D=-0.7$. 
 b)  Loss distribution in  $[0,T]$,  $T=1$,  of system  (\ref{OurBanks12}), (\ref{OurBanksCI12}), (\ref{rho})  when $N=10$,  $\epsilon=0.05$, $\alpha=50$, $\gamma=-50$ (solid line) and  loss distribution in  $[0,T]$,  $T=1$,  of   system  (\ref{OurBanks12}), (\ref{OurBanksCI12}), (\ref{rho})  when $N=10$, $\epsilon=0$,   $\alpha=\gamma=0$ (dashed  line).}

 \label{fig6}
  \end{figure}
  \begin{figure}
  \centerline{\includegraphics[height=6cm]{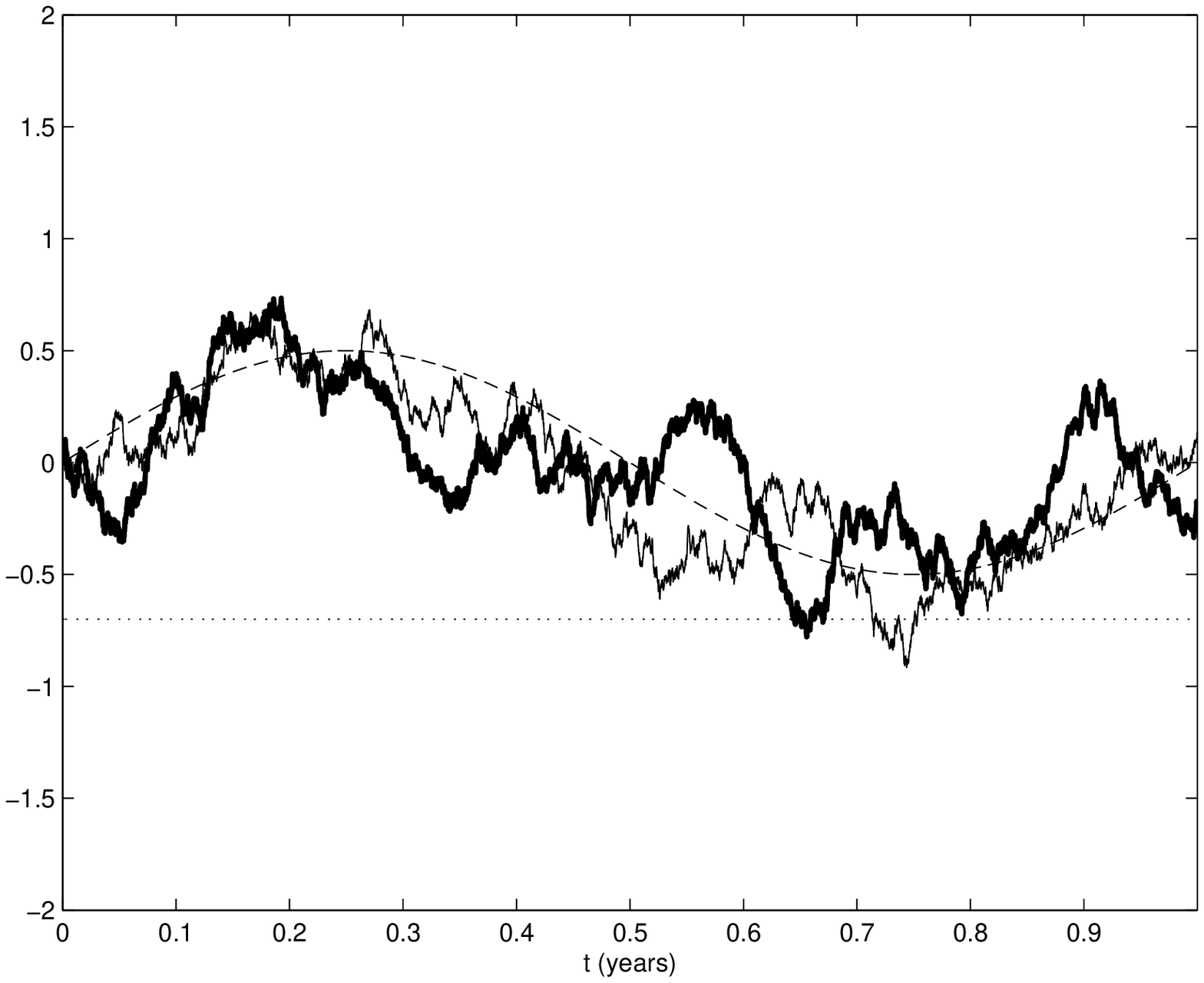}a)\includegraphics[height=6cm]{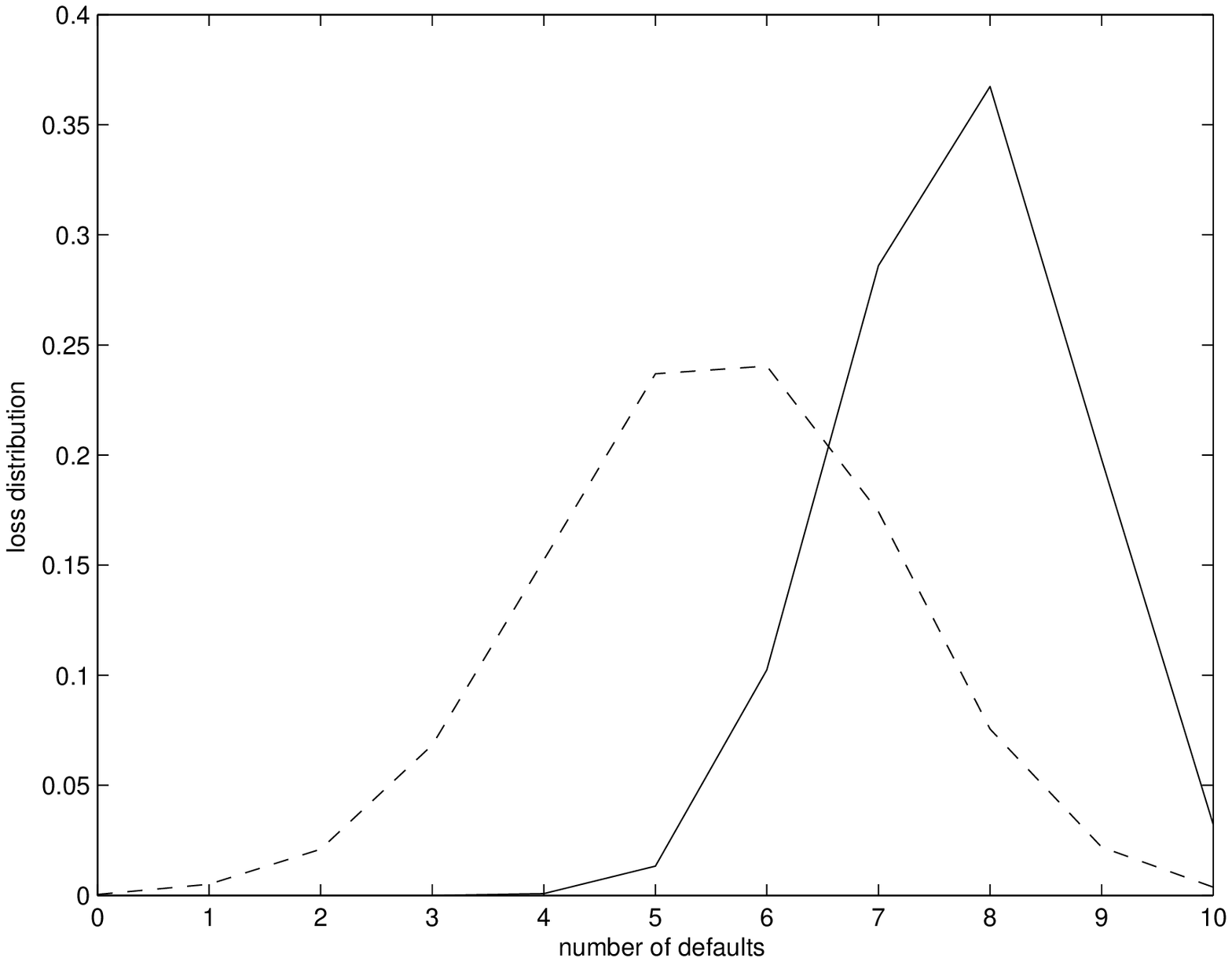}b)}
  \caption{\footnotesize 
  a) The solid line shows  one  trajectory of   $X_{t}^{1}$,  $ t \in [0,T]$, $T=1$,   solution of  system (\ref{OurBanks12}), (\ref{OurBanksCI12}), (\ref{rho}) when $N=10$,  $\epsilon=0.05$, $\alpha=10,$  $\gamma=-100$.  
   The dashed line shows  the target  trajectory  $\xi(t)=\displaystyle 0.5 \sin (2 \pi t)$, $ t \in [0,T]$, $T=1$.
 The thick solid line   shows one trajectory  of the  solution ${\cal X}_{t}$,  $ t \in [0,T]$, $T=1$,  of the corresponding   mean field initial value problem  (\ref{12OurMF1}), (\ref{12OurMF2}), (\ref{12OurMF3}). 
 The dotted  line shows the default level $D=-0.7$. 
 b)  Loss distribution in  $[0,T]$,  $T=1$,  of system  (\ref{OurBanks12}), (\ref{OurBanksCI12}), (\ref{rho})  when $N=10$, $\epsilon=0.05$,  $\alpha=10$, $\gamma=-100$ (solid line) and  loss distribution in  $[0,T]$,  $T=1$,  of   system  (\ref{OurBanks12}), (\ref{OurBanksCI12}), (\ref{rho})  when $N=10$, $\epsilon=0$,  $\alpha=\gamma=0$ (dashed  line).}

 \label{fig7}
 \end{figure}


%


\section{An optimal control problem for the mean-field equations}\label{sec4}
%


Let $ \R^{+} $ be  the set of positive real numbers.  Note  that both the mean field equation (\ref{FMF}) of the banking  system model (\ref{FBanks}),   (\ref{FBanksCI}), (\ref{rho})  and  the mean field equation  (\ref{12OurMF1}) of the  banking system model (\ref{OurBanks12}),   (\ref{OurBanksCI12}), (\ref{rho})  are  of  the   form:
\begin{equation}\label{GeneralMF}
 d{\cal Z}_{t}= \beta (t,{\cal Z}_{t}) dt+ \sigma dW_{t},\,  t>0, 
\end{equation}
where the function  $ \beta:  \R^{+} \times \R \rightarrow \R$  is a sufficiently regular  function.  In  fact the choices ${\cal Z}_{t}={\cal Y}_{t},  $   $ t\ge0,$ and
\begin{equation}\label{beta1}
\beta=\beta_{t}=\beta(t,{\cal Y}_{t})=-\alpha {\cal Y}_{t}, \quad t>0,   
\end{equation}
reduce (\ref{GeneralMF}) to  (\ref{FMF}),  similarly  the choices  ${\cal Z}_{t}={\cal X}_{t},  $   $ t\ge0,$ and
 \begin{equation}\label{beta2}
\beta=\beta_{t}= \beta(t,{\cal X}_{t})=
\alpha \left(   \mediaX_{t} - {\cal X}_{t} \right) + \gamma \left(   \mediaX_{t} - \xi_{t}^{-} \right)  +(\xi_{t}^{+})' ,\quad t>0, 
\end{equation}
where $\displaystyle (\xi_{t}^{+})' =  \frac{d\xi_{t}^{+}}{dt}$, $ t\ge0$,  and    $\mediaX_{t}$, $t\ge 0$,  is defined  by   (\ref{12OurMF2}), (\ref{12OurMF3}) reduce (\ref{GeneralMF}) to  (\ref{12OurMF1}). Finally  note that  the choices  ${\cal Z}_{t}={\cal X}_{t},  $   $ t\ge0,$ and
 \begin{equation}\label{beta2bis}
\beta=\beta_{t}= \beta(t,{\cal X}_{t})=
\alpha \left(   \mediaX_{t} - {\cal X}_{t} \right) + \gamma \left(   \mediaX_{t} - \xi_{t} \right)  +(\xi_{t})' ,\quad t>0, 
\end{equation}
where $ \mediaX_{t}$,    $ t\ge0,$ is defined  by   (\ref{OurMF2}),  (\ref{OurMF3})  reduce  (\ref{GeneralMF}) to   (\ref{OurMF1}).
However, as already seen in Section \ref{sec3},   equations   (\ref{OurMF2}),  (\ref{OurMF3})    imply $ \mediaX_{t}=\xi_{t}$,    $ t\ge0,$ and  this means that in (\ref{beta2bis}) $\gamma$ is multiplied by zero.

In     (\ref{FMF})  and in (\ref{12OurMF1})  $\alpha$ and $\gamma$ are    constants.  In this Section  the parameters   $\alpha$ and $\gamma$  are   functions of time.

Let  $\br$ be the set of the real square integrable processes defined in  $[0,T_{1}]$, $T_{1}>0$. A real stochastic process $\zeta=\zeta_{t}, $ $t\in [0,T_{1}]$, belongs to $\br$ if and only if $\displaystyle  \E \left(  \int_{0}^{T_{1}}   \zeta_{t}^{2} \, dt \right) <  +\infty$.

Given  the constants  $\lambda>0$,  $T_{1}>0$,  and the target   trajectory     $\xi_{t}$, $ t\in [0,T_{1}]$,  let us determine  
$\beta=\beta_{t} \in  \br$  as   solution of the  following   stochastic optimal  control   problem:
\begin{eqnarray}
&&  \min_{\beta \in \br}  U_{\lambda} (\beta) , 
  \label{Func1}
  \end{eqnarray}
  where
  \begin{eqnarray}
    U_{\lambda} (\beta) =   \E     \left( \int_{0}^{T_{1}} \left[\left({\cal Z}_{t}-\xi_{t} \right)^{2} +\lambda \beta_{t}^{2}\right] dt  \right) , \, \, \, \beta=\beta_{t}, \, \, t\in [0,T_{1}], \, \,   \lambda>0, 
  \label{Func2}
  \end{eqnarray}
  subject to the constraints:
   \begin{eqnarray} 
&&        d{\cal Z}_{t}   =\beta \, dt +\sigma dW_{t},    \quad   t\in [0,T_{1}],     \label{Func3}\\
&&        {\cal Z}_{0}   =\xi_{0}^{+}.       \label{Func4}
\end{eqnarray}
In the  control problem   (\ref{Func1}), (\ref{Func2}), (\ref{Func3}), (\ref{Func4}) the function $ U_{\lambda} (\beta)$, $ \beta=\beta_{t}, $ $t\in [0,T_{1}], $  $  \lambda>0,  $  is the utility function and  $\beta=\beta_{t}$, $t\in [0,T_{1}], $ is the control variable. 
 
The  utility function $ U_{\lambda} (\beta)$, $ \beta\in \br, $   $  \lambda>0,   $  defined in   (\ref{Func2})  is the sum of two terms. The first term,    $\displaystyle \E      \left( \int_{0}^{T_{1}} \left({\cal Z}_{t}-\xi_{t} \right)^{2} dt \right), $  penalizes the departure of ${\cal Z}_{t}$,  $t\in [0,T_{1}]$,   from the target  trajectory $\xi_{t}$, $t\in [0,T_{1}]$.
The  second one,   $\displaystyle \E      \left( \int_{0}^{T_{1}} \lambda \beta_{t} ^{2} \,  dt  \right) $,  penalizes   the  ``size'' of the  control variable $\beta_{t}$, $t\in [0,T_{1}]$.

We   solve  problem  (\ref{Func1}), (\ref{Func2}), (\ref{Func3}), (\ref{Func4}) using   the  dynamic programming principle (see \cite{Kolosov}).    That is let 
\begin{eqnarray}
&&V(t,{\cal Z})=
 \min_{\beta_{t} \in \br}  \E     \left( \int_{t}^{T_{1}} \left[\left({\cal Z}_{\tau}-\xi_{\tau} \right)^{2} +\lambda \beta_{\tau} ^{2}\right] d\tau   \Big|  {\cal Z}_{t}={\cal Z}\right) , \nonumber
\\
&&     \hskip7truecm  {\cal Z} \in \R,  \, t\in [0,T_{1}],   \label{valuefunction}
\end{eqnarray}
 be the value function  of the  control problem   (\ref{Func1}), (\ref{Func2}), (\ref{Func3}), (\ref{Func4}).
 The  function $V(t,{\cal Z})$, ${\cal Z} \in \R, $ $  t\in [0,T_{1}]$, 
 satisfies the following Hamilton, Jacobi, Bellman equation (see \cite{Kolosov}):
\begin{eqnarray}
&& \frac{\partial } {\partial t} V(t,{\cal Z}) +\frac12 \sigma^{2}   \frac{\partial^{2} } {\partial {\cal Z}^{2}} V(t,{\cal Z}) +
  \calH\left( \frac{\partial } {\partial {\cal Z}} V(t,{\cal Z}) \right)+\left({\cal Z}-\xi_{t} \right)^{2} =0 ,
\nonumber  \\
&&     \hskip7truecm  {\cal Z} \in \R,  \, t\in [0,T_{1}],    \label{HJB}
\end{eqnarray}
with  final condition:
\begin{eqnarray}
&& V(T_{1},{\cal Z})=0 , \quad  {\cal Z} \in \R, 
  \label{HJBCI}  
\end{eqnarray}
where
\begin{eqnarray}  \label{Hamilt} 
&& \calH(p)= \min_{\delta \in \R} \left( \delta p    + \lambda  \delta^{2}\right)=-\frac{p^{2}}{4 \lambda},  \quad p \in \R, 
\end{eqnarray}
is the Hamiltonian function  of the optimal control problem   (\ref{Func1}), (\ref{Func2}), (\ref{Func3}), (\ref{Func4}).\\
Using    (\ref{Hamilt})  equation  (\ref{HJB}) becomes:
\begin{eqnarray}
&& \frac{\partial } {\partial t} V(t,{\cal Z}) +\frac12 \sigma^{2}   \frac{\partial^{2} } {\partial {\cal Z}^{2}} V(t,{\cal Z}) -
\frac{1}{4 \lambda}\left( \frac{\partial } {\partial {\cal Z}} V(t,{\cal Z}) \right)^{2}
+\left({\cal Z}-\xi_{t} \right)^{2} =0 ,
\nonumber \\
&&     \hskip7truecm  {\cal Z} \in \R,  \, t\in [0,T_{1}],   \label{HJB2} 
\end{eqnarray}
with the final condition (\ref{HJBCI}).

From the knowledge  of the value function $V$ solution of   (\ref{HJB2}),  (\ref{HJBCI}) we determine  the optimal control  $\beta_{t}$,  $  t\in [0,T_{1}], $  solution of (\ref{Func1}), (\ref{Func2}), (\ref{Func3}), (\ref{Func4})
using the  condition:
\begin{eqnarray} \label{soluz} 
&& 
\beta=\beta_{t}=\beta(t,{\cal Z}_{t})= - \frac{1}{2\lambda} \frac{\partial } {\partial {\cal Z}} V(t,{\cal Z})\Big|_{{\cal Z}={\cal Z}_{t}}, \quad  t\in [0,T_{1}] , 
 \end{eqnarray}
where ${\cal Z}_{t}$, $t\in [0,T_{1}]$,  is the solution of (\ref{Func3}), (\ref{Func4}) when $\beta$=$\beta_{t}$,  $  t\in [0,T_{1}], $   is given by  (\ref{soluz}).

Proposition  1   reduces the solution of     (\ref{HJB2}), (\ref{HJBCI}) to the solution of a final value problem  for a system of  ordinary differential equations.

\noindent
{ \bf Proposition 1.} A solution of problem  (\ref{HJB2}),  (\ref{HJBCI})  is given by:
\begin{eqnarray} \label{soluzProp} 
&& 
V(t,{\cal Z})=a(t) +b(t) {\cal Z}  +c(t) {\cal Z}^{2}, \quad {\cal Z} \in \R, \, \,  t\in [0,T_{1}] ,
 \end{eqnarray}
where  the functions $a(t)$,  $b(t)$, $c(t)$, $  t\in [0,T_{1}],$  are  solution of   the following   final value problem for a  system of  Riccati  ordinary differential equations:
\begin{eqnarray}
&&    \frac{\partial } {\partial t} a= -\sigma^{2} c + \frac{b^{2} } {4 \lambda} - \xi_{t}^{2}, \quad   \quad  t\in [0,T_{1}] , \quad   \quad \quad a(T_{1})=0,  \label{Riccati1}  \\[2mm]
&&    \frac{\partial } {\partial t} b=  \frac{b c} { \lambda} +2 \xi_{t}, \quad   \quad  t\in [0,T_{1}] , \quad   \quad \quad b(T_{1})=0,  \label{Riccati2}
  \\[2mm]
&&    \frac{\partial } {\partial t} c=  \frac{c^{2} } { \lambda} - 1, \quad   \quad  t\in [0,T_{1}] , \quad   \quad \quad c(T_{1})=0. \label{Riccati3}  
 \end{eqnarray}
The optimal control  $\beta=\beta_{t}$, $t\in [0,T_{1}]$,  solution  of  problem (\ref{Func1}), (\ref{Func2}), (\ref{Func3}), (\ref{Func4})  determined by  the value function (\ref{soluzProp})    is:
\begin{eqnarray} \label{soluzPropBeta} 
\beta=\beta_{t}=\beta(t,{\cal Z}_{t})= - \frac{1}{2\lambda} \left( b(t)+ 2 c(t)  {\cal Z}_{t} \right), \quad   t\in [0,T_{1}] . 
 \end{eqnarray}

\noindent
{\bf Proof.}  Let $V(t,{\cal Z})$, ${\cal Z} \in \R, $ $  t\in [0,T_{1}], $  be   of the form   (\ref{soluzProp}) (see \cite{Kolosov}),  
substituting  (\ref{soluzProp})  in  (\ref{HJB2}),  (\ref{HJBCI}) and  using the  polynomial identity principle   it is easy to see  that  problem (\ref{HJB2}),  (\ref{HJBCI}) reduces to   the final value problem  (\ref{Riccati1}), (\ref{Riccati2}), (\ref{Riccati3}). Equation (\ref{soluzPropBeta}) 
for the optimal control of problem (\ref{Func1}), (\ref{Func2}), (\ref{Func3}), (\ref{Func4})   follows from  (\ref{soluzProp}),  (\ref{soluz}).\\

Note that the constraint  (\ref{Func3}) of  the  optimal  control   problem   (\ref{Func1}), (\ref{Func2}), (\ref{Func3}), (\ref{Func4}) is a model equation  that can  represent     the mean field equations  (\ref{FMF}) or  (\ref{12OurMF1})  when we have respectively ${\cal Z}_{t}={\cal Y}_{t},  $   $ t\ge0,$ or ${\cal Z}_{t}={\cal X}_{t},  $   $ t\ge0,$  and   equations  (\ref{beta1})  or (\ref{beta2}) hold.
Problem  (\ref{Func1}), (\ref{Func2}), (\ref{Func3}), (\ref{Func4}) is the  optimal control problem for the mean field approximation  (\ref{12OurMF1}), (\ref{12OurMF2}), (\ref{12OurMF3})     of  (\ref{OurBanks12}),   (\ref{OurBanksCI12}), (\ref{rho})   used  to control the behaviour  of the trajectories of the  banking system model (\ref{OurBanks12}),   (\ref{OurBanksCI12}), (\ref{rho}).
In fact  we determine  the  functions  $\alpha_{t}$, $\gamma_{t}$, $  t\in [0,T_{1}], $   that must be substituted  in     (\ref{OurBanks12}),   (\ref{OurBanksCI12}), (\ref{rho})  to force the trajectories of (\ref{OurBanks12}),   (\ref{OurBanksCI12}), (\ref{rho})  to swarm  around 
$\xi_{t}$,  $t\in [0,T_{1}]$, from  the  optimal control  (\ref{soluzPropBeta})  of  problem (\ref{Func1}), (\ref{Func2}), (\ref{Func3}), (\ref{Func4})   imposing    ${\cal Z}_{t}={\cal X}_{t},  $   $t\in [0,T_{1}]$,  and    (\ref{beta2}).
%
%
%
Comparing  the optimal  control (\ref{soluzPropBeta}) with equation   (\ref{beta2})  when ${\cal Z}_{t}={\cal X}_{t},  $  $t\in [0,T_{1}]$,   and  using the  polynomial identity principle we have:
\begin{eqnarray} 
&&\alpha_{t}= \frac{c(t)}{\lambda} ,    \quad   t\in [0,T_{1}],  \label{alphaott} \\[2mm]
&&\gamma_{t}=\frac{1}{  \mediaX_{t} - \xi_{t}^{-} } \left( - \frac{c(t)}{\lambda}   \mediaX_{t}  -
 \frac{1}{2\lambda} b(t)-(\xi_{t}^{+})'\right) ,    \quad   t\in [0,T_{1}].  \label{gammaott}
 \end{eqnarray}

Note   that  the function $ \mediaX_{t}$, $t\in [0,T_{1}]$, is defined  by the equations  (\ref{12OurMF2}), (\ref{12OurMF3})  and  contains  an integral that depends from  $\gamma_{t}$,  $t\in [0,T_{1}]$.  This means  that equation (\ref{gammaott}) is an integral  equation in the unknown $\gamma_{t}$,  $t\in [0,T_{1}]$.
The conditions  $\alpha=\alpha_{t} \ge 0, $ $ \gamma=\gamma_{t} \le 0$,  $t\in [0,T_{1}]$, imposed in 
(\ref{alphagammatime}) are  constraints that must be satisfied. When   they are not  satisfied by the choices  of $\alpha_{t},$ $ \gamma_{t}$,  $t\in [0,T_{1}]$, made in (\ref{alphaott}),  (\ref{gammaott}) they must be enforced. 
Let us point out   that in the numerical experiments discussed  in Section  \ref{sec5}  the function  $\alpha_{t}$,  $t\in [0,T_{1}]$,  determined by  (\ref{alphaott}) is  always positive  while the function  $\gamma_{t}$,  $t\in [0,T_{1}]$,  determined by  (\ref{gammaott}) is  not always smaller or equal than  zero. 
When  the function  $\gamma_{t}$ determined by  (\ref{gammaott}) is positive,  we enforce the choice  $\gamma_{t}=0$ to guarantee (\ref{alphagammatime}). 
More in general  the enforcement of the constraints (\ref{alphagammatime})   can be done solving   the equations (\ref{alphaott}),  (\ref{gammaott}) in the   least squares sense after imposing  (\ref{alphagammatime}) as a constraint.
The choice of  $\alpha_{t}$, $\gamma_{t}$,  $t\in [0,T_{1}]$,  that follows from   (\ref{alphaott}),  (\ref{gammaott}), (\ref{alphagammatime}) induces a swarming effect of the trajectories of (\ref{OurBanks12}),   (\ref{OurBanksCI12}), (\ref{rho}) around the target trajectory $\xi_{t}$,  $t\in [0,T_{1}]$, similar to the swarming effect induced by the optimal control $\beta$ soultion of  (\ref{Func1}), (\ref{Func2}), (\ref{Func3}), (\ref{Func4})  on the trajectories of   (\ref{Func3}), (\ref{Func4}).

Note that the polynomial identity principle  can be used when we compare (\ref{soluzPropBeta})   with    (\ref{beta2}) 
 but cannot be used if we compare  (\ref{soluzPropBeta}) with   (\ref{beta1})   or  with    (\ref{beta2bis}).  
In this last case this is due to the fact that in      (\ref{beta2bis}) $\gamma$ is multiplied by zero. 
In fact in (\ref{beta1}) and   (\ref{beta2bis}) $\beta$ is a polynomial of degree one  respectively in ${\cal Y}_{t}$ and  in ${\cal X}_{t}$  with only one  coefficient that can be chosen. 
This is the reason  to  consider  model 
(\ref{OurBanks12}),   (\ref{OurBanksCI12}), (\ref{rho}) instead of model (\ref{OurBanks}),   (\ref{OurBanksCI}), (\ref{rho}).

Note that  proceeding as done previously    a possibly high dimensional dynamical  system (i.e. the banking system model  (\ref{OurBanks12}),   (\ref{OurBanksCI12}), (\ref{rho}))  is   controlled  using a  law  (i.e. the law  (\ref{alphaott}),  (\ref{gammaott}), (\ref{alphagammatime})) deduced from the solution  of a control problem (i.e.  problem (\ref{Func1}), (\ref{Func2}), (\ref{Func3}), (\ref{Func4})) for  a low dimensional dynamical  system (i.e. the  mean field equations (\ref{12OurMF1}), (\ref{12OurMF2}), (\ref{12OurMF3}) of  the banking system model  (\ref{OurBanks12}),   (\ref{OurBanksCI12}), (\ref{rho})). 
  
 The  possibility, exploited here,   of governing a   high dimensional dynamical  system using its mean field approximation,   is based  on    the fact that  in the   problem  considered the population of agents  (i.e.  the population of banks) described by the high dimensional dynamical   system  is homogeneous (i.e. the banks are all equal and in the limit $N$ goes to infinity all the banks behave like the mean bank). This  feature of the banking system model (\ref{OurBanks12}),   (\ref{OurBanksCI12}), (\ref{rho})  is   a rather  common  feature   that is  found  in many other circumstances in science and engineering. 
 
%
\section{Systemic risk governance}\label{sec5}
%
%
%
Let $ 0\le \tau_{1}<\tau_{2} < +\infty$  we     consider  in the  model (\ref{OurBanks12}), (\ref{OurBanksCI12}), (\ref{rho}) the  problem of  governing    the probability    of   systemic risk in the bounded time interval $[\tau_{1},\tau_{2}]$. 
The systemic  risk governance   is   based on   the choice  of the target trajectory  $\xi_{t}$, $t\in  [\tau_{1},\tau_{2}]$,  and  of   the parameter $\epsilon>0$. The governance   exploits   the solution of the control problem  (\ref{Func1}), (\ref{Func2}), (\ref{Func3}), (\ref{Func4}), and   its relation with 
 the banking system  model   (\ref{OurBanks12}), (\ref{OurBanksCI12}), (\ref{rho}) when the functions $\alpha_{t}$, $\gamma_{t}$, $  t\in  [\tau_{1},\tau_{2}], $  are chosen  as suggested by   (\ref{alphaott}),  (\ref{gammaott}), (\ref{alphagammatime}).
  The goal of  the governance is to keep        the   probability of    systemic risk in the time interval $[\tau_{1},\tau_{2}]$,  $\pr (SR_{[\tau_{1},\tau_{2}]})$,  between two given thresholds. 
Given $\epsilon$   this goal is  pursued choosing  at  time $t=\tau_{1}$   the target trajectory $\xi_{t}$, $t\in [\tau_{1},\tau_{2}]$.
    In fact  we know that  increasing    $\xi_{t}-D>0$,   $t\in [\tau_{1},\tau_{2}]$,  the systemic risk  probability in $ [\tau_{1},\tau_{2}]$ decreases  and  that  decreasing   $\xi_{t}-D>0$, $t\in [\tau_{1},\tau_{2}]$,  the systemic risk  probability in $[\tau_{1},\tau_{2}]$ increases. 
 
 Going into details  given  $\epsilon$,    the thresholds $S_{1}$, $S_{2}$ such that  $0<S_{1}<S_{2}<1$ and  $\xi_{\tau_{1}}$   we want
 to determine a   target   trajectory    $\xi_{t}$, $t\in [\tau_{1},\tau_{2}]$,  such  that     the probability of    systemic risk  in the time interval $[\tau_{1},\tau_{2}]$  satisfies the following inequalities:
\begin{eqnarray} \label{striscia}
S_{1} \le \pr (SR_{[\tau_{1},\tau_{2}]}) \le S_{2}.
 \end{eqnarray}
%
%
%
Let us define some simple rules  that can be  used  to choose $\xi_{t}$, $t\in [\tau_{1},\tau_{2}]$.  At time $t=\tau_{1}$    the  ``simplest''      choice   for  the target   trajectory in the time interval $[\tau_{1},\tau_{2}]$  is   $\xi_{t}=\xi_{\tau_{1}}$,  $t\in [\tau_{1},\tau_{2}]$.   In correspondence to this choice  the monetary authority   determines the functions   $\alpha_{t}$, $\gamma_{t}$, $t\in [\tau_{1},\tau_{2}]$,   using  (\ref{alphaott}),  (\ref{gammaott}), (\ref{alphagammatime})
and  evaluates the probability   of    systemic risk   in the time interval $[\tau_{1},\tau_{2}]$ associated to the choices of 
$\xi_{t}$, $\alpha_{t}$, $\gamma_{t}$, $t\in [\tau_{1},\tau_{2}]$ done,  that is evaluates  $\pr (SR_{[\tau_{1},\tau_{2}]})$.
Note that $\pr (SR_{[\tau_{1},\tau_{2}]})$ depends  not only from  $\xi_{t}$, $\alpha_{t}$, $\gamma_{t}$, $t\in [\tau_{1},\tau_{2}]$, but   also  from the values taken by the variables $X_{\tau_{1}}^{i}$, $  i=1,2,\ldots,N$.
Based on the value of $\pr (SR_{[\tau_{1},\tau_{2}]})$   the monetary authority acts as follows:
 \begin{enumerate} 
 \item[] {\it Strategy 1}:  if $\pr (SR_{[\tau_{1},\tau_{2}]}) < S_{1}$  the monetary authority    changes the  choice of  target   trajectory  $\xi_{t}$, $t\in [\tau_{1}, \tau_{2}]$,   to ``swarm''  the trajectories   of the  banking system model (\ref{OurBanks12}), (\ref{OurBanksCI12}), (\ref{rho})   ``downward''  (i.e.  decreases  $\xi_{t}-D>0$, $t\in [\tau_{1}, \tau_{2}]$);
 \item[] {\it Strategy 2}:   if $\pr (SR_{[\tau_{1},\tau_{2}]}) > S_{2}$  the monetary authority    changes the  choice of  target   trajectory  $\xi_{t}$, $t\in [\tau_{1}, \tau_{2}]$,   to ``swarm''  the trajectories   of the  banking system model (\ref{OurBanks12}), (\ref{OurBanksCI12}), (\ref{rho}) ``upward'' (i.e.  increases   $\xi_{t}-D>0$, $t\in [\tau_{1}, \tau_{2}]$);
 \item[] {\it Strategy 3}:  if  $S_{1} \le \pr (SR_{[\tau_{1},\tau_{2}]}) \le S_{2} $    the monetary authority    leaves the target   trajectory chosen, that is  $\xi_{t}=\xi_{\tau_{1}}$, $t\in [\tau_{1}, \tau_{2}]$,   unchanged. 
  \end{enumerate} 
 %


 Let us  discuss   some numerical experiments  that  illustrate  a simple implementation of the previous strategies.
    In  the   numerical experiments  we present the simulation of the governance of the  systemic risk   in the next year during a period of two years. The governance  takes place quarterly  and consists,  at the beginning of each quarter,  in choosing    the target trajectory of the    log-monetary reserves  of the ``ideal bank'' in  the next year     and in translating this choice  in rules for the banks determining  the values  of the functions  $\alpha_{t}$,  $\gamma_{t}$  in the next year using (\ref{alphaott}),  (\ref{gammaott}),  (\ref{alphagammatime}). These  choices are done pursuing the goal of  enforcing   the probability of systemic risk  in the next  year to satisfy  the inequalities  (\ref{striscia}). 
For simplicity we restrict the choice of the target trajectory  to a finite set of functions   constructed with some simple rules.

  In the first numerical experiment  we study  the governance of  the  probability of systemic risk  in the next year  when there are no shocks acting on the banking system.
In the second and in the third numerical experiment  we study  the governance of  the probability of systemic risk  in the next year  in presence of positive and negative shocks acting on the banking system.
These shocks are represented as abrupt changes of $\sigma$, that is we consider $\sigma=\sigma_{t}$, $t>0$, to be a piecewise constant function. 

{\it Numerical experiment 1.}
We choose a time horizon $T_{2}$ of three years, that is $T_{2}=3$,  and the time step  of  the governance decisions    $\Delta \tau=1/4$, that is we take  systemic risk   governance   decisions  quarterly.
In the time interval $[0,T_{2}]$ we consider   the time intervals $ [\tau_{1}^{j},\tau_{2}^{j}] \subset [0,T_{2}]$, $T_{2}=3$,   where $\displaystyle \tau_{1}^{j}=j \cdot \Delta \tau$ and $\tau_{2}^{j}=\tau_{1}^{j}+1$, $j=0,1,\ldots,8$. 
The remaining  parameters  of the model  used in the numerical example   are: $N=10$, $\xi_{0}=1$, $\epsilon=0.1$, $\lambda=0.001$,  $S_{1}= 0.03$, $S_{2}=  0.05$,  $D=0.3$,   $\sigma=1$.
Note that in this experiment  we have $\sigma=\sigma_{t}=1$, $t\in[0,T_{2}]$, $T_{2}=3$. 
 The choice  of constant  volatility corresponds  to the fact that there are no  shocks acting on the banking system during the time interval considered.
In   the numerical experiments the number $N=10$ of banks is constant  during the time evolution, that is  there are no bank defaults 
during the time evolution. 

 In each time interval $ [\tau_{1}^{j},\tau_{2}^{j}]$, $j=0,1,\ldots,8,$  the model   (\ref{OurBanks12}),   (\ref{OurBanksCI12}), (\ref{rho}) reduces to:
\begin{eqnarray}
&& dX_{t}^{i}= \frac{\alpha}{N}  \sum_{j=1}^{N} \left( X_{t}^{j} -X_{t}^{i}\right) dt+ 
                          \gamma \left( \frac{1}{N} \sum_{j=1}^{N} X_{t}^{j} -\xi_{t}^{-}\right)  dt +d\xi_{t}^{+} + \sigma dW_{t}^{i},\nonumber\\
&& \hskip4truecm   t\in [\tau_{1}^{j},\tau_{2}^{j}], \,  i=1,2,\ldots,N ,  \, \,  j=0,1,\ldots,8, \label{OurBanks12interv}
\end{eqnarray}
with initial condition:
\begin{eqnarray}
    X_{\tau_{1}^{0}}^{i}=\xi_{0}^{+}   , \, \,  i=1,2,\ldots,N,   \quad X_{\tau_{1}^{j}}^{i} =X_{\tau_{2}^{j-1}}^{i}, \, \,  i=1,2,\ldots,N,   \, \,  j=1,2,\ldots,8,       \label{OurBanksCI12interv}
\end{eqnarray}
 and 
\begin{eqnarray}\label{rhointerv}
\E(dW_{t}^{i} dW_{t}^{k})&=&\delta_{i,k} dt,\,\, t\in [\tau_{1}^{j},\tau_{2}^{j}], \, \,  i,k=1,2,\ldots,N,  \, \,  j=0,1,\ldots,8.
\end{eqnarray}

As  explained   in Section \ref{sec2},  for $j=0,1,\ldots,8$, the probability  of    systemic risk  in the  time interval $ [\tau_{1}^{j},\tau_{2}^{j}]$ in the model (\ref{OurBanks12interv}),   (\ref{OurBanksCI12interv}), (\ref{rhointerv})  is evaluated using  statistical simulation starting from $10^{4}$    trajectories of   (\ref{OurBanks12interv}),   (\ref{OurBanksCI12interv}), (\ref{rhointerv}) generated numerically. These trajectories are  obtained   integrating     (\ref{OurBanks12interv}),   (\ref{OurBanksCI12interv}), (\ref{rhointerv}) using  the explicit Euler method with  time step $\Delta t=10^{-4}$.

For $j=0,1,\ldots,8$ in order to keep  the probability  of systemic risk   in the time  interval $ [\tau_{1}^{j},\tau_{2}^{j}]$    between the thresholds  $S_{1}= 0.03$ and $S_{2}= 0.05$, we give to the monetary authority   eight choices of  target trajectory    that  push  the trajectories   of  the $j$-th   model  (\ref{OurBanks12interv}),   (\ref{OurBanksCI12interv}), (\ref{rhointerv})  ``downward'' and  eight  choices of  target trajectory  that  push  the trajectories   of  the $j$-th  model (\ref{OurBanks12interv}),   (\ref{OurBanksCI12interv}), (\ref{rhointerv})   ``upward''.  
Finally the monetary authority   has the possibility of choosing  the target trajectory   constant during  the next year   and  equal to   its  value     at the decision time. This last choice leaves the trajectories of the $j$-th model (\ref{OurBanks12interv}),   (\ref{OurBanksCI12interv}), (\ref{rhointerv}) ``unchanged'', that is  it does not push them  ``upward'' or  ``downward''.
   The piecewise linear functions listed in (\ref{possibility})  are the  target trajectories used to implement these   choices.  

 Initially, for simplicity,  starting   from  $\xi_{0}=1$, $\epsilon=0.1$,  the monetary authority  chooses  $\xi_{t}=\xi_{0}^{+}=\xi_{0}+\epsilon=1.1$,  $t\in [\tau_{1}^{0},\tau_{2}^{0}]=[0,1]$ as  target trajectory.
For $j=0,1,\ldots,8$, given the time interval   $ [\tau_{1}^{j},\tau_{2}^{j}]$,  the monetary authority assumes      $\xi_{t}=\xi_{\tau_{1}^{j}}$,  $t \in  [\tau_{1}^{j},\tau_{2}^{j}]$,     and  evaluates $\pr (SR_{[\tau_{1}^{j},\tau_{2}^{j}]})$.
Depending  from  the value of $\pr (SR_{[\tau_{1}^{j},\tau_{2}^{j}]})$ the monetary authority acts   as suggested by 
  {\it Strategy 1}-{\it Strategy 3}. This last step  is done choosing    one of the following target trajectories:
\begin{eqnarray} \label{possibility}
\nonumber P_{n}:  \quad \quad 
\xi_{t}=\xi_{t,n}=
\left\{%
\begin{array}{ll}
  \displaystyle \frac{n}{8} \,  (t-\tau_{1}^{j})+\xi_{\tau_{1}^{j}} ,  & t\in [\tau_{1}^{j},\tau_{1}^{j} + \Delta \tau]  , \\[4mm]
  \displaystyle  \frac{n}{8}  \, \Delta t+\xi_{\tau_{1}^{j}} ,  & t\in (\tau_{1}^{j} + \Delta \tau, \tau_{2}^{j}]  , 
\end{array}%
\right.\\[4mm]
 \quad  n=-8,-7,..,0,..,7,8.
 \end{eqnarray}
Note that in (\ref{possibility}) when  $0<n\le 8$ (respectively $-8\le n<0$)  the target trajectory is a  non decreasing  (respectively non increasing) piecewise linear function.   When $n=0$  the target trajectory is  constant. 
Recall that we are always assuming that $\xi_{t}-D>0$, $t\in   [\tau_{1}^{j},\tau_{2}^{j}]$, and that choices of      target trajectories 
defined  in (\ref{possibility})   that violate this constraint are not considered. 
 That is   when the monetary authority  chooses a  
target trajectory $P_{n}$ with $0<n\le 8$ in the time interval   $ [\tau_{1}^{j},\tau_{2}^{j}]$  the trajectories   of the  banking system model (\ref{OurBanks12interv}),   (\ref{OurBanksCI12interv}), (\ref{rhointerv})   move ``upward'' and the probability  of systemic risk in the next year decreases,  vice versa when the monetary authority  chooses a target trajectory  $P_{n}$ with $-8\le n<0$  in the time interval   $ [\tau_{1}^{j},\tau_{2}^{j}]$  the trajectories   of the  banking system model (\ref{OurBanks12interv}),   (\ref{OurBanksCI12interv}), (\ref{rhointerv})   move ``downward''  and the probability  of systemic risk in the next year increases. 
Finally  when  the monetary authority  chooses the target trajectory   $P_{0}$ in the time interval   $ [\tau_{1}^{j},\tau_{2}^{j}]$,     the  character of the trajectories of the  banking system model (\ref{OurBanks12interv}),   (\ref{OurBanksCI12interv}), (\ref{rhointerv}) is left unchanged  and the corresponding probability  of systemic risk in the next year is left approximately  unchanged.  
%

Note that the  choices made    in (\ref{possibility}) are  only illustrative. Many   other choices of target trajectories   are possible to  implement  {\it Strategy 1}-{\it Strategy 3}.

In Figures 8-10 we show the governance of   model (\ref{OurBanks12interv}),   (\ref{OurBanksCI12interv}), (\ref{rhointerv})  
 made by the monetary authority  at  the beginning of each quarter pursuing the goal of keeping    the probability  of    systemic risk in  the next year  between the thresholds  $S_{1}= 0.03$ and $S_{2}= 0.05$. 
In Figures 8-10   the  choices of target trajectories   considered at  the beginning of each quarter by the monetary authority in the decision making process  are shown.  The   target trajectory   
marked with a dot at its endpoint  is the final choice of  target trajectory made  by the monetary authority.
The monetary authority runs through the  possible choices  of  target trajectories listed in  (\ref{possibility})   in their natural order  according to    {\it Strategy 1}-{\it Strategy 3}  starting from the choice    $P_{0}$  and evaluates the probability of systemic risk in the next year associated to each target trajectory  considered   before making its final decision (see Figures 8-10). The first target trajectory encountered that 
has   the associated probability  of systemic risk in the next year that satisfies (\ref{striscia})  is chosen. 
Figure 11 shows   the probability of   systemic risk in the next year evaluated  at  the beginning of each quarter during the time interval  $[0,2]$  both  in presence  and in  absence of governance. 
 In presence   of governance the probability shown is the probability of systemic risk in the next year evaluated  at  the beginning of each quarter when  the target trajectory chosen is the final choice made by the monetary authority in that quarter   marked with a dot at its endpoint  in  Figures 8-10.
In absence of governance we choose $\xi_{t}=\xi_{0}=1$,  $\alpha_{t}=20$,  $\gamma_{t}=-1$, $t\in [0,T_{2}]$, $T_{2}=3$,  $\epsilon=0.1$,  and we evaluate
the probability of   systemic risk in the next year   at  the beginning of each  quarter. 
The     choices   made in absence of governance guarantee that   the probability  of systemic risk   in the next year  at time $t=0$  is between the   given thresholds.

Recall that in the  Numerical experiment 1 we have $\sigma=\sigma_{t}=1$, $ t \in  [0,T_{2}]$, $T_{2}=3$.
Figures 8-10  show that, when the volatility  $\sigma$ is constant   in the time interval $ [0,T_{2}]$, $T_{2}=3$,  the systemic risk  governance  of   the monetary authority   is reduced to the choice of   the target trajectory at the beginning of the time interval considered, that is at time $t=0$.  When this choice is done correctly   continuing with a constant target trajectory    or with small variations of it   in each successive time interval   $ [\tau_{1}^{j},\tau_{2}^{j}]$,  $j=0,1,\ldots,8$,   is  sufficient to  keep   the probability  of systemic risk  in the next year  between the   given thresholds.    Figure 11 shows  that,
when the volatility  $\sigma$ is constant in the time interval  $ [0,T_{2}]$, $T_{2}=3$,  the presence or absence of governance does not make a significant difference provided that  in absence of governance  a good choice of the target trajectory and of the  values of the parameter $\alpha$ and  $\gamma$ is done at time $t=0$. 
  
 In the  next experiments  we consider two examples  of non constant  volatility  in the time interval  $ [0,T_{2}]$,  $T_{2}=3$, that is   we consider the case when  there are shocks acting on the banking system during the time interval considered. 

{\it Numerical experiments 2 and 3.}
The setting of the Numerical experiments 2 and 3 is  the  same    one  of      Numerical experiment 1   except for the fact that in these   two experiments   the volatility $\sigma$ is  not constant.   We allow   abrupt changes  of  volatility to model shocks acting on the banking system. \\
In  Numerical experiment 2   we choose  the volatility   $\sigma$  as follows:
\begin{eqnarray} \label{sigmatime1}
\sigma=\sigma_{1,t}=
\left\{%
\begin{array}{ll}
  \displaystyle 1 ,  & t\in [0,1]  , \\
  \displaystyle  1.5 ,  & t\in (1,3].
\end{array}%
\right.
 \end{eqnarray}
In  Numerical experiment 3 we choose  the volatility   $\sigma$   as follows:
\begin{eqnarray} \label{sigmatime2}
\sigma=\sigma_{2,t}=
\left\{%
\begin{array}{ll}
  \displaystyle 1 ,  & t\in [0,0.8]  , \\
  \displaystyle  0.3 ,  & t\in (0.8, 1.2],\\ 
   \displaystyle  1.3 ,  & t\in (1.2, 3] .
\end{array}%
\right.
 \end{eqnarray}
The function   (\ref{sigmatime1})  models  a positive  volatility shock, while the function   (\ref{sigmatime2}) models a negative volatility shock   followed by a positive one.

Note that in the Numerical experiments 2 and 3 at  the beginning of each quarter the monetary authority   in the process of   making   its decision about systemic risk governance assumes that the volatility  in the next year  is constant at  its present value. That is the monetary authority  does not foresee the  volatility shocks, simply reacts to them when they occur.

Figures 12-14 show the governance of the banking system model (\ref{OurBanks12interv}),   (\ref{OurBanksCI12interv}), (\ref{rhointerv})  in the time interval $ [\tau_{1}^{j},\tau_{2}^{j}]$, $j=0,1,\ldots,8$,  made by the monetary authority  at  the beginning of each quarter in order to keep   the probability  of    systemic risk  during  the next year  between the thresholds  $S_{1}= 0.03$ and $S_{2}= 0.05$ when the volatility  $\sigma$  is  given by  (\ref{sigmatime1}). 
Figures 16-18 show the same quantities  of Figures 12-14 when the volatility  $\sigma$ is given by  (\ref{sigmatime2}). 
In particular in each image of  Figures  12-14 and  of Figures  16-18  we show the choices   of the target trajectories for the next year  considered  by the monetary authority at  the beginning of each quarter  and the final choice of the target trajectory  for the next year  made    at the beginning of each  quarter (marked with a dot at its endpoint). 
%
%
Figure 15 and  19  show  respectively in  the time interval  $[0,2]$ the probability  of   systemic risk in the next year evaluated  at  the beginning of each quarter  both  in presence  and in  absence of governance   for    the Numerical experiments 2 and 3. In absence of governance  as done in Numerical experiment 1 we have chosen:  $\xi_{t}=1$,  $\alpha_{t}=20$,   $\gamma_{t}=-1$,  $t\in [0,T_{2}]$,  $T_{2}=3$, $\epsilon=0.1$.

Figures 12-19 show that,  as expected,   the governance of the systemic risk in presence of positive  and/or negative  volatility shocks is more demanding than in absence of volatility  shocks.
 Nevertheless  Figures 12-19   show that the elementary  systemic risk  governance  outlined   in this Section  after a few adjustments  is able  to keep 
  the probability   of    systemic risk   in the next year between the    assigned thresholds.
 %



%
%
%
%
 \begin{figure}
 \centerline{\includegraphics[height=13cm]{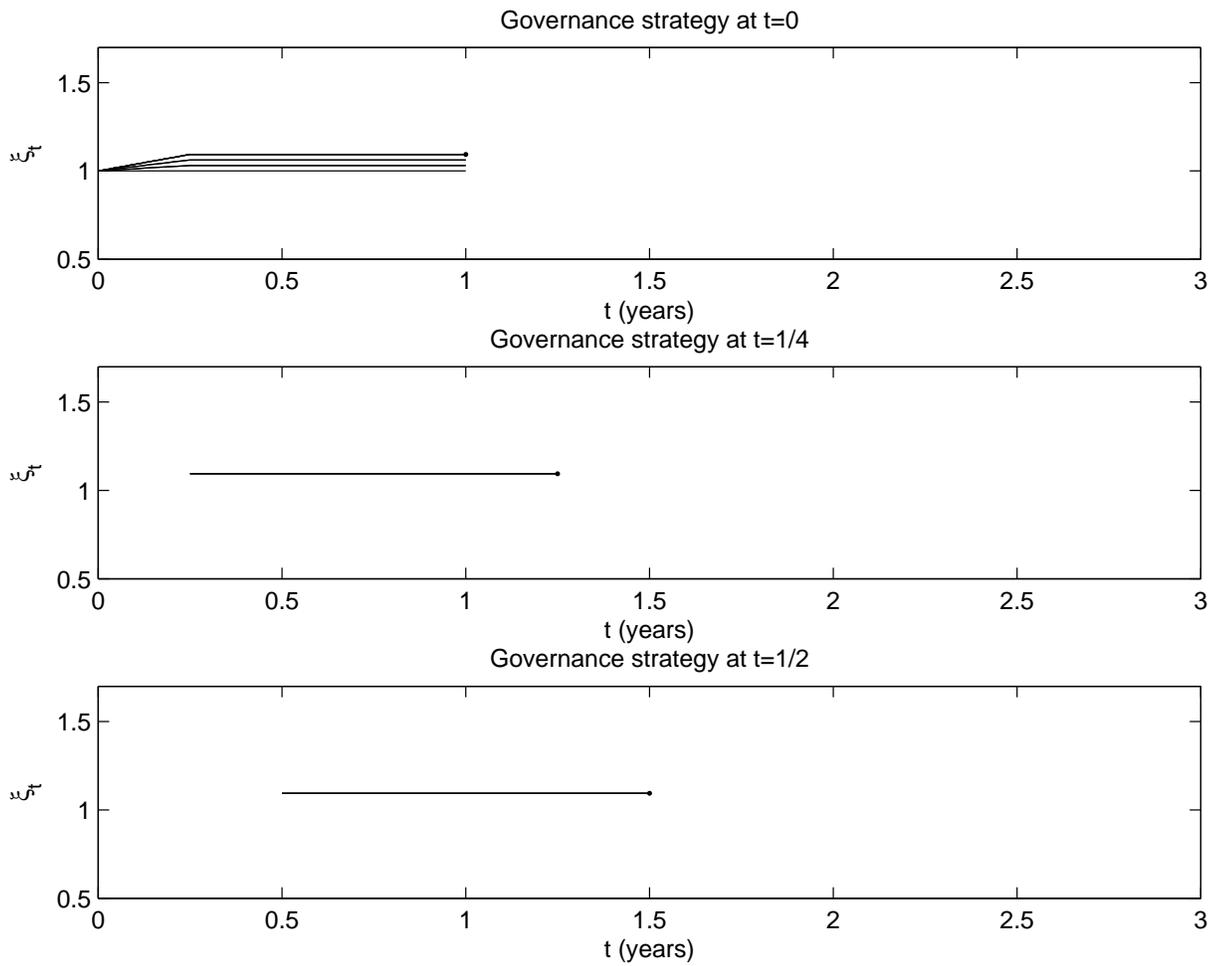}}
  \caption{\footnotesize  Numerical experiment 1:  choices of target trajectory  considered by the monetary authority at the beginning of a quarter; the target trajectory   marked with a   dot at its endpoint  is  the final choice of  the monetary authority.}
 \label{fig8}
  \end{figure}
    \begin{figure}
 \centerline{\includegraphics[height=13cm]{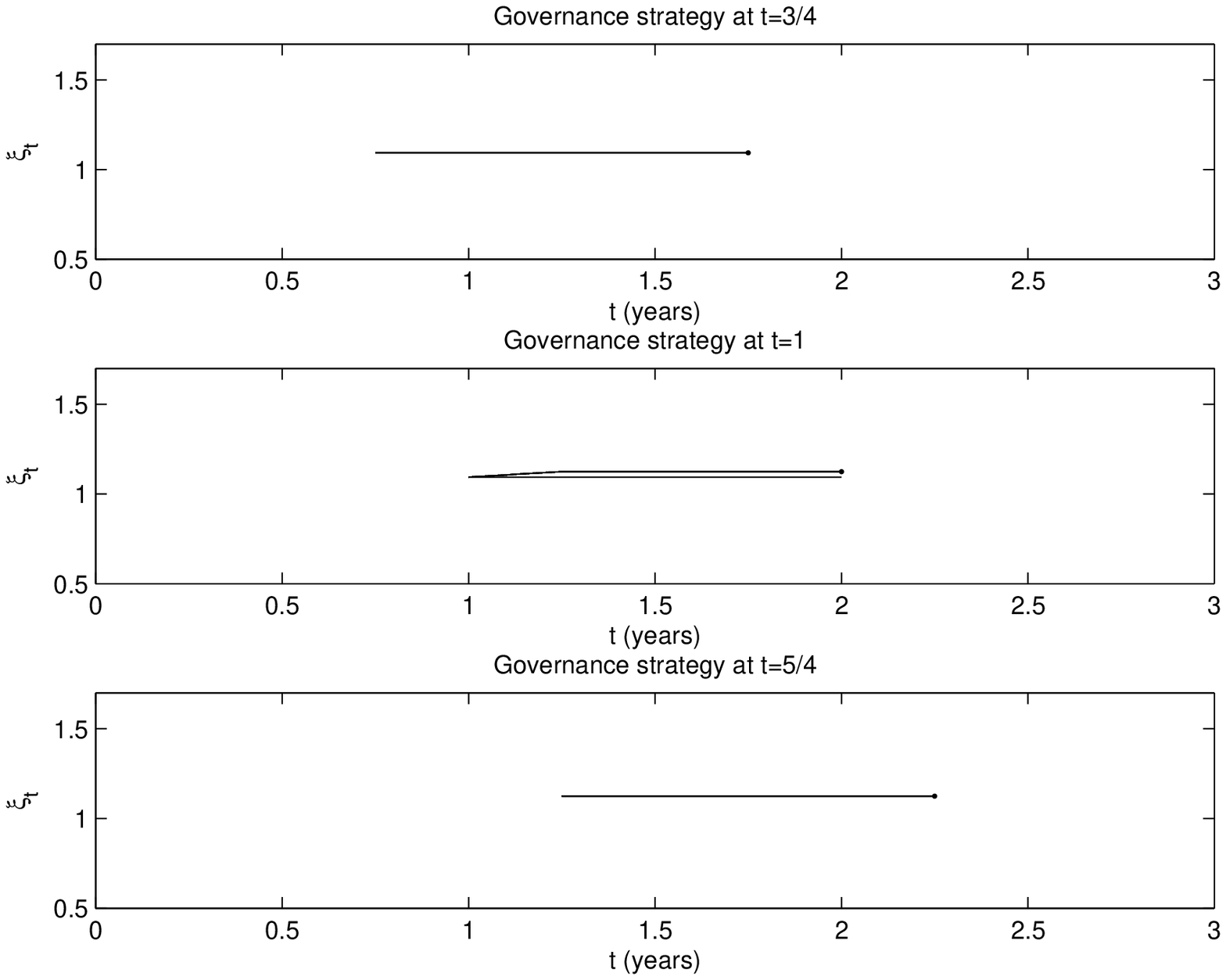}}
   \caption{\footnotesize  Numerical experiment 1:  choices of target trajectory  considered by the monetary authority at the beginning of a quarter;  the target trajectory   marked with a   dot at its endpoint  is  the final choice of  the monetary authority.}
 \label{fig9}
  \end{figure}
  \begin{figure}
 \centerline{\includegraphics[height=13cm]{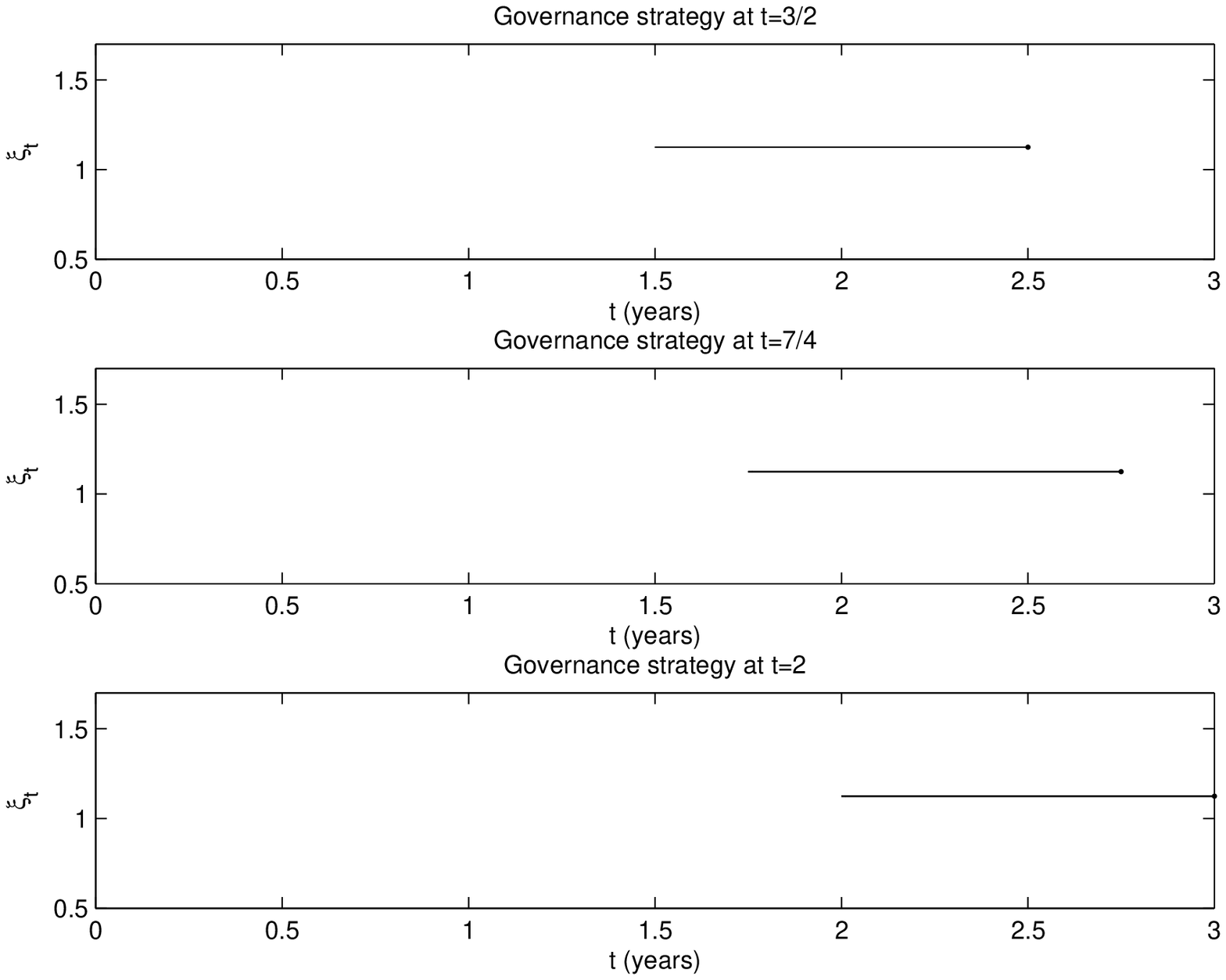}}
  \caption{\footnotesize  Numerical experiment 1: choices of target trajectory  considered by the monetary authority at the beginning of a quarter;  the target trajectory   marked with a   dot at its endpoint  is  the final choice of  the monetary authority.}
 \label{fig10}
  \end{figure}
  \begin{figure}
 \centerline{\includegraphics[height=8cm]{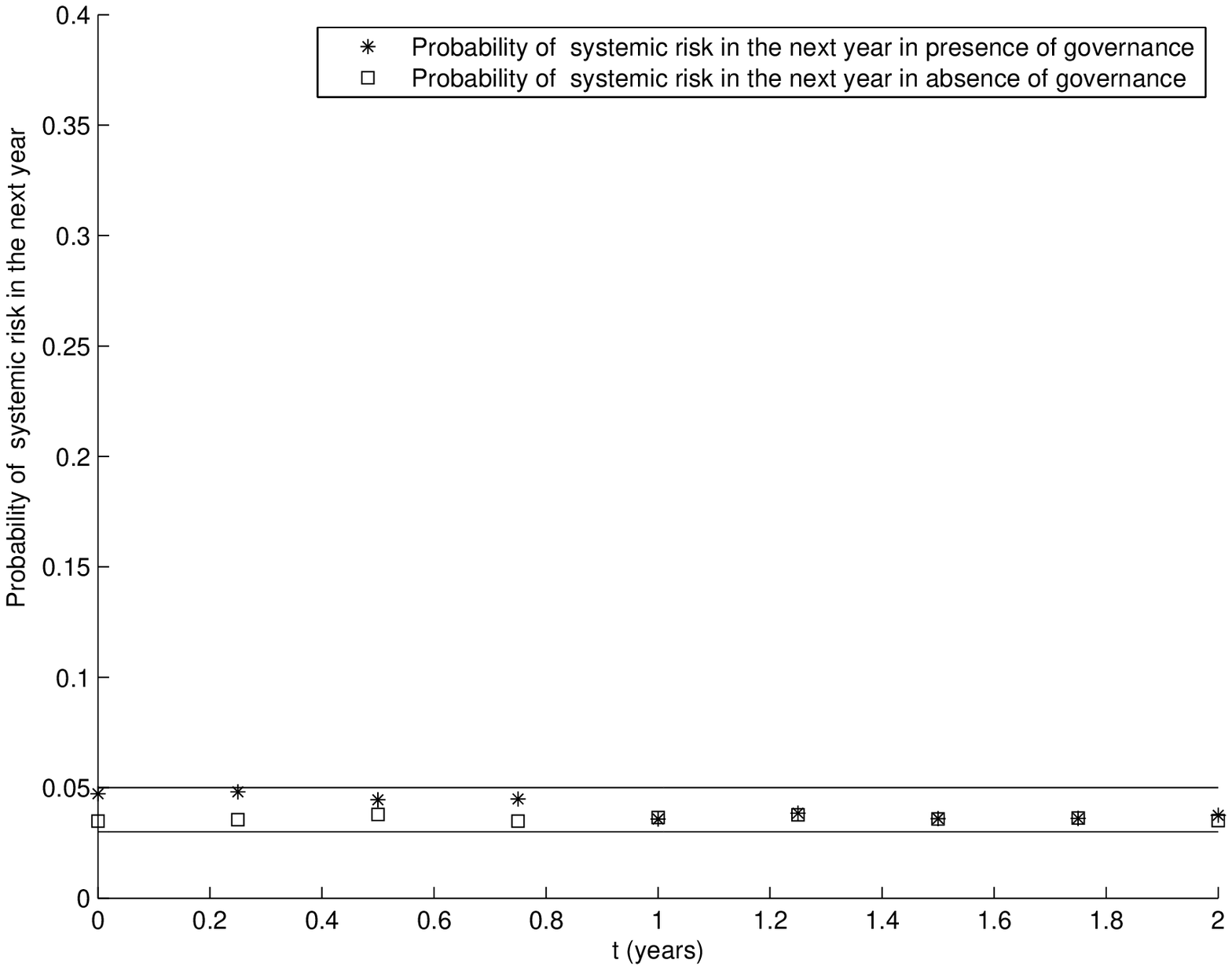}}
  \caption{\footnotesize    Numerical experiment 1: probability of   systemic risk in the next year evaluated  at  the beginning of each quarter 
  in the time interval  $ [0,T_{2}]= [0,2]$   in presence of governance ($*$) and in  absence of governance ($\square$).
  }
 \label{fig11}
  \end{figure}
%
%
 \begin{figure}
 \centerline{\includegraphics[height=13cm]{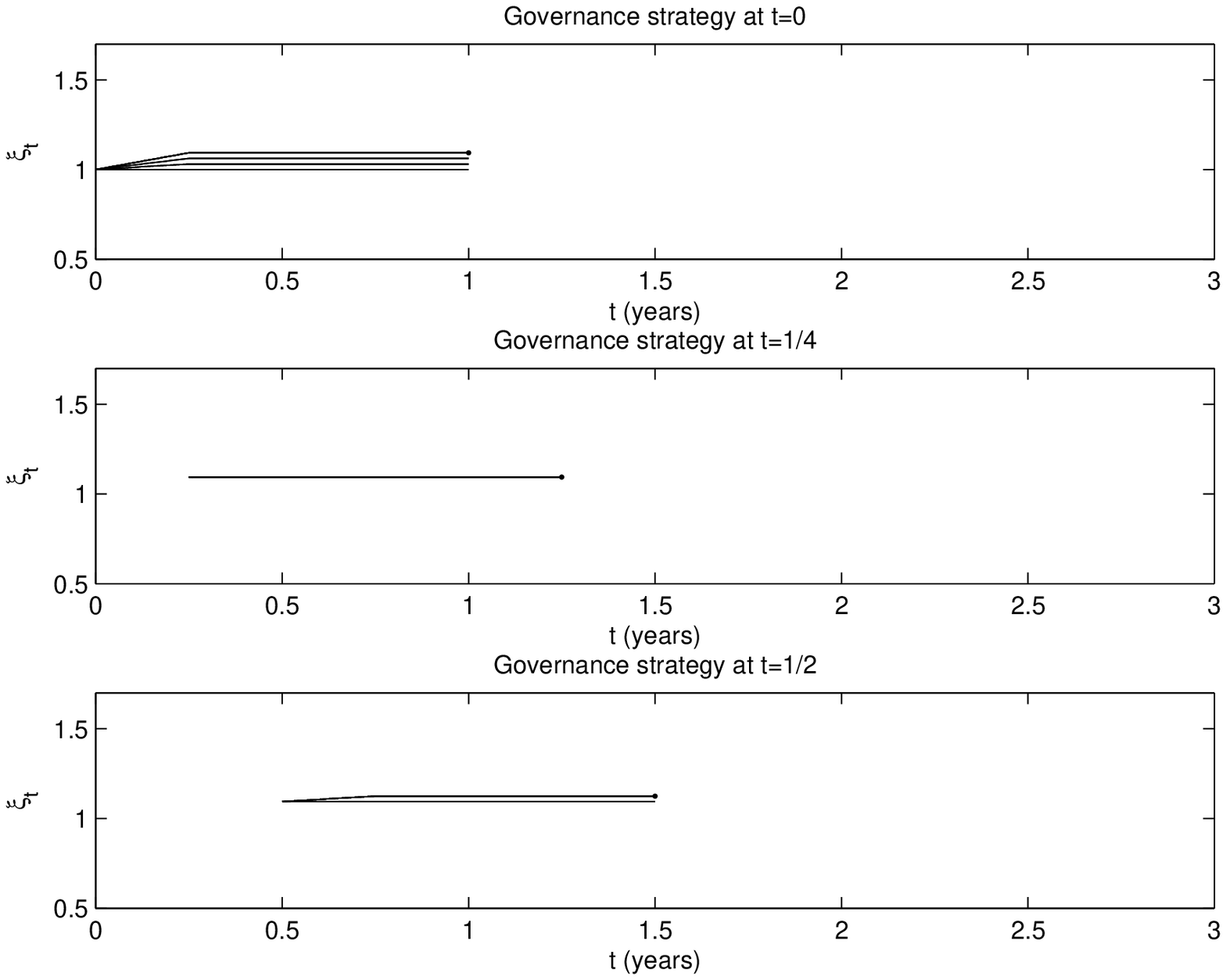}}
  \caption{\footnotesize  Numerical experiment 2:    choices of target trajectory  considered by the monetary authority at the beginning of a quarter;   the target trajectory   marked with a   dot at its endpoint  is  the final choice of  the monetary authority.}
 \label{fig12}
  \end{figure}
   \begin{figure}
 \centerline{\includegraphics[height=13cm]{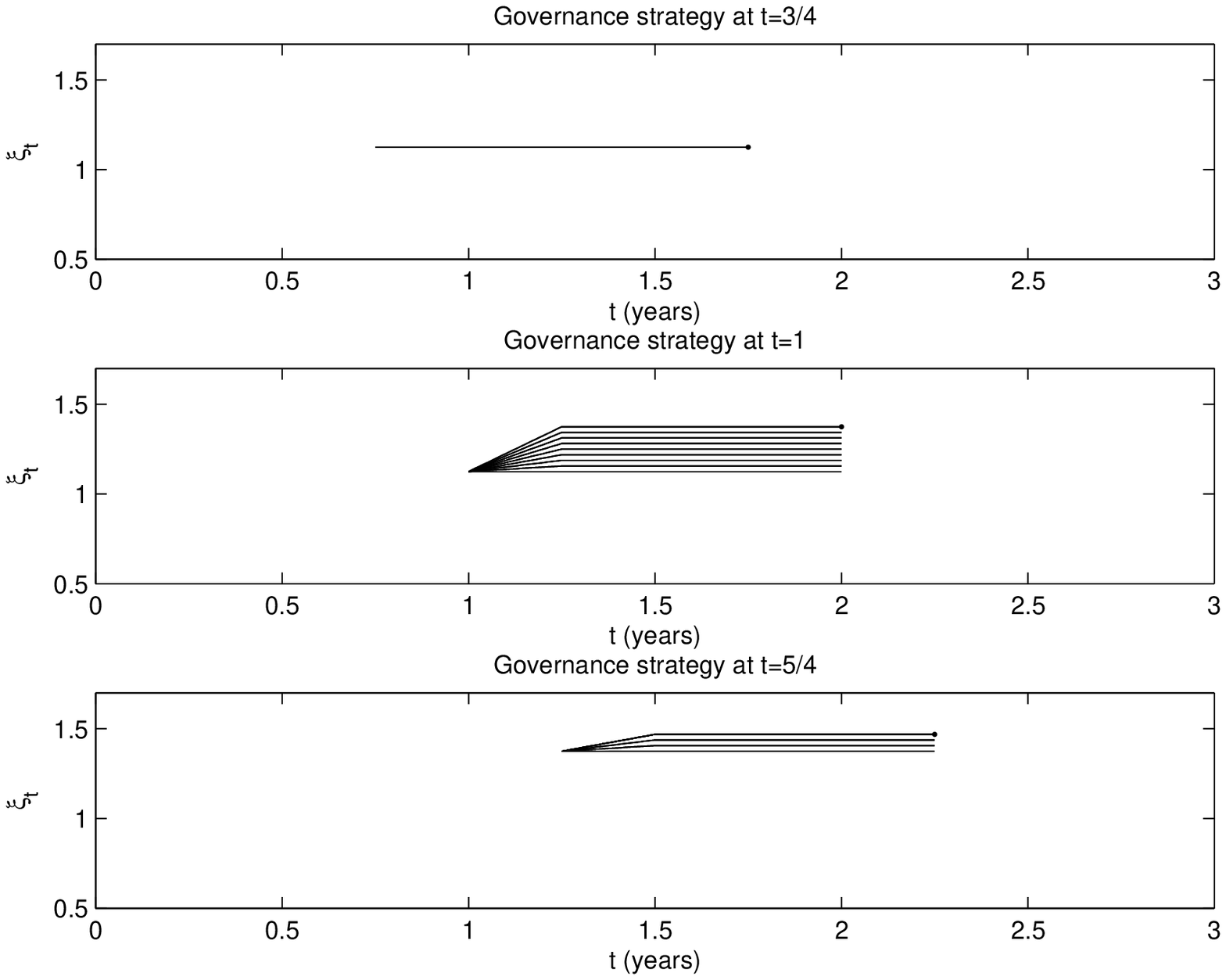}}
  \caption{\footnotesize  Numerical experiment 2: choices of target trajectory  considered by the monetary authority at the beginning of a quarter;  the target trajectory   marked with a   dot at its endpoint  is  the final choice of  the monetary authority.} 
  \label{fig13}
  \end{figure}
  \begin{figure}
 \centerline{\includegraphics[height=13cm]{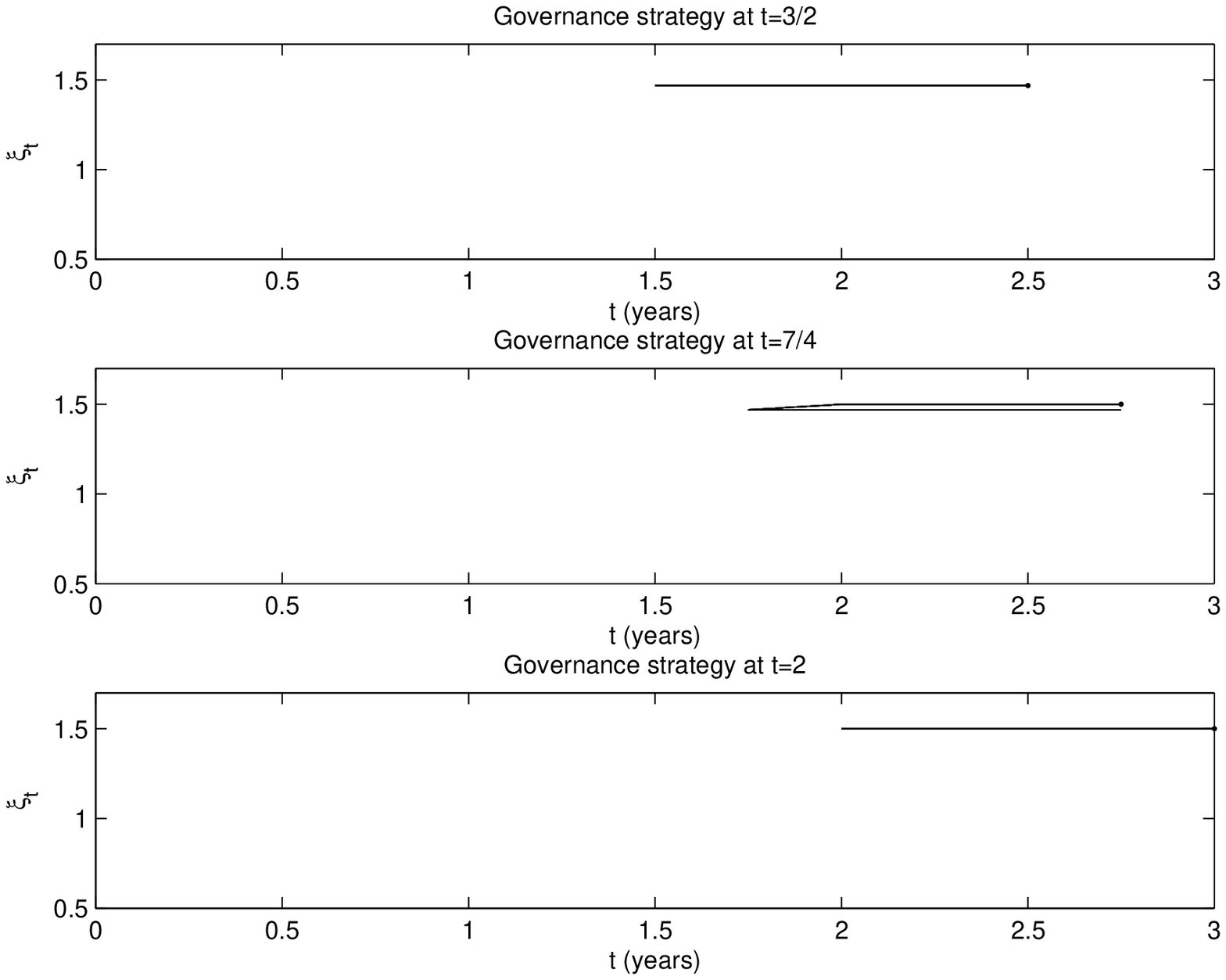}}
  \caption{\footnotesize  Numerical experiment 2:    choices of target trajectory  considered by the monetary authority at the beginning of a quarter; the target trajectory   marked with a   dot at its endpoint  is  the final choice of  the monetary authority.}
 \label{fig14}
  \end{figure}
  \begin{figure}
 \centerline{\includegraphics[height=8cm]{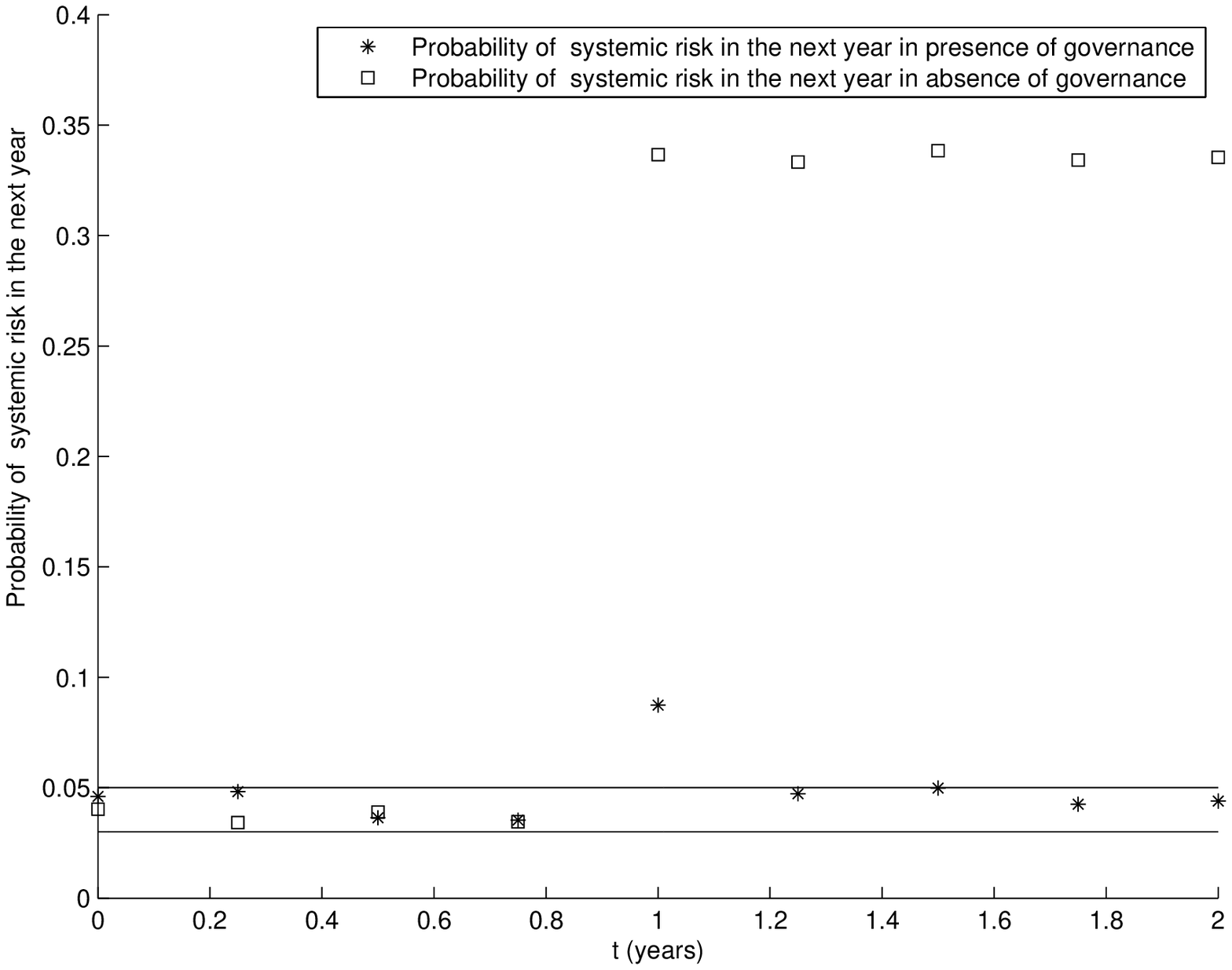}}
  \caption{\footnotesize    Numerical experiment 2: probability of   systemic risk in the next year evaluated  at  the beginning of each quarter 
  in the time interval  $ [0,T_{2}]= [0,2]$   in presence of governance ($*$) and in  absence of governance ($\square$).
}
 \label{fig15}
  \end{figure}
 %
%
%

 \begin{figure}
 \centerline{\includegraphics[height=13cm]{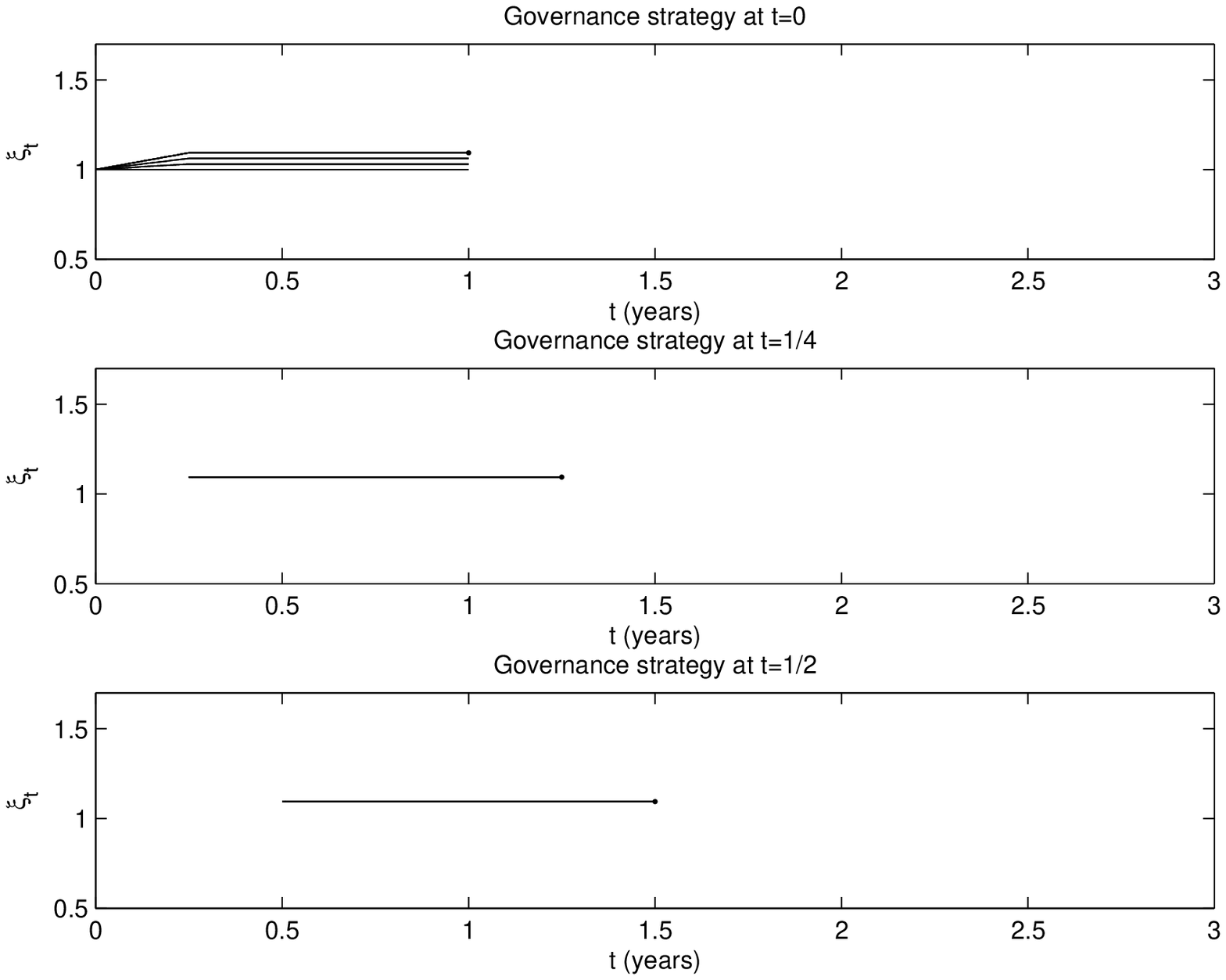}}
  \caption{\footnotesize  Numerical experiment 3:  choices of target trajectory  considered by the monetary authority at the beginning of a quarter; the target trajectory   marked with a   dot at its endpoint  is  the final choice of  the monetary authority.}
 \label{fig16}
  \end{figure}
   \begin{figure}
 \centerline{\includegraphics[height=13cm]{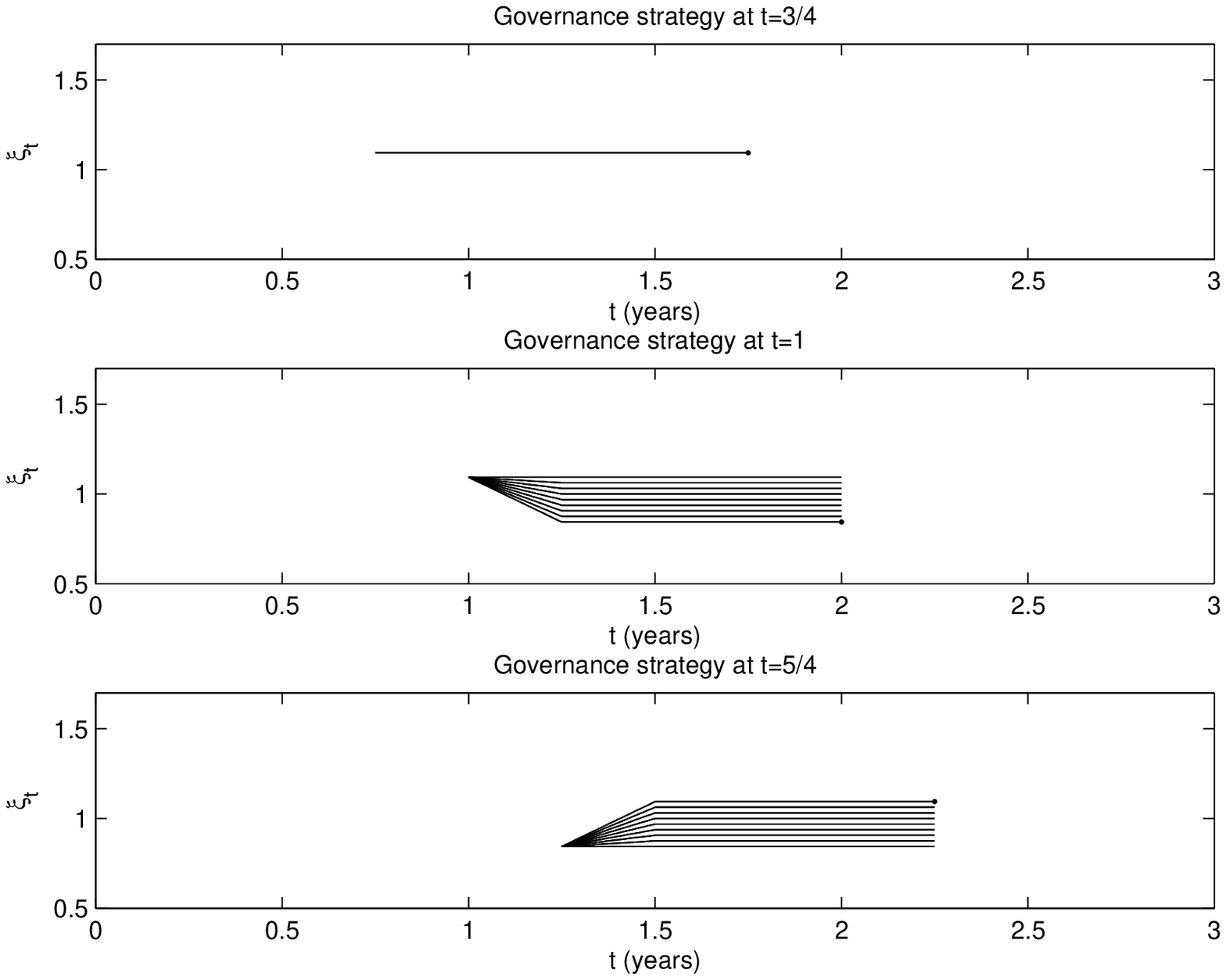}}
  \caption{\footnotesize  Numerical experiment 3:  choices of target trajectory  considered by the monetary authority at the beginning of a quarter;  the target trajectory   marked with a   dot at its endpoint  is  the final choice of  the monetary authority.}
 \label{fig17}
  \end{figure}
  \begin{figure}
 \centerline{\includegraphics[height=13cm]{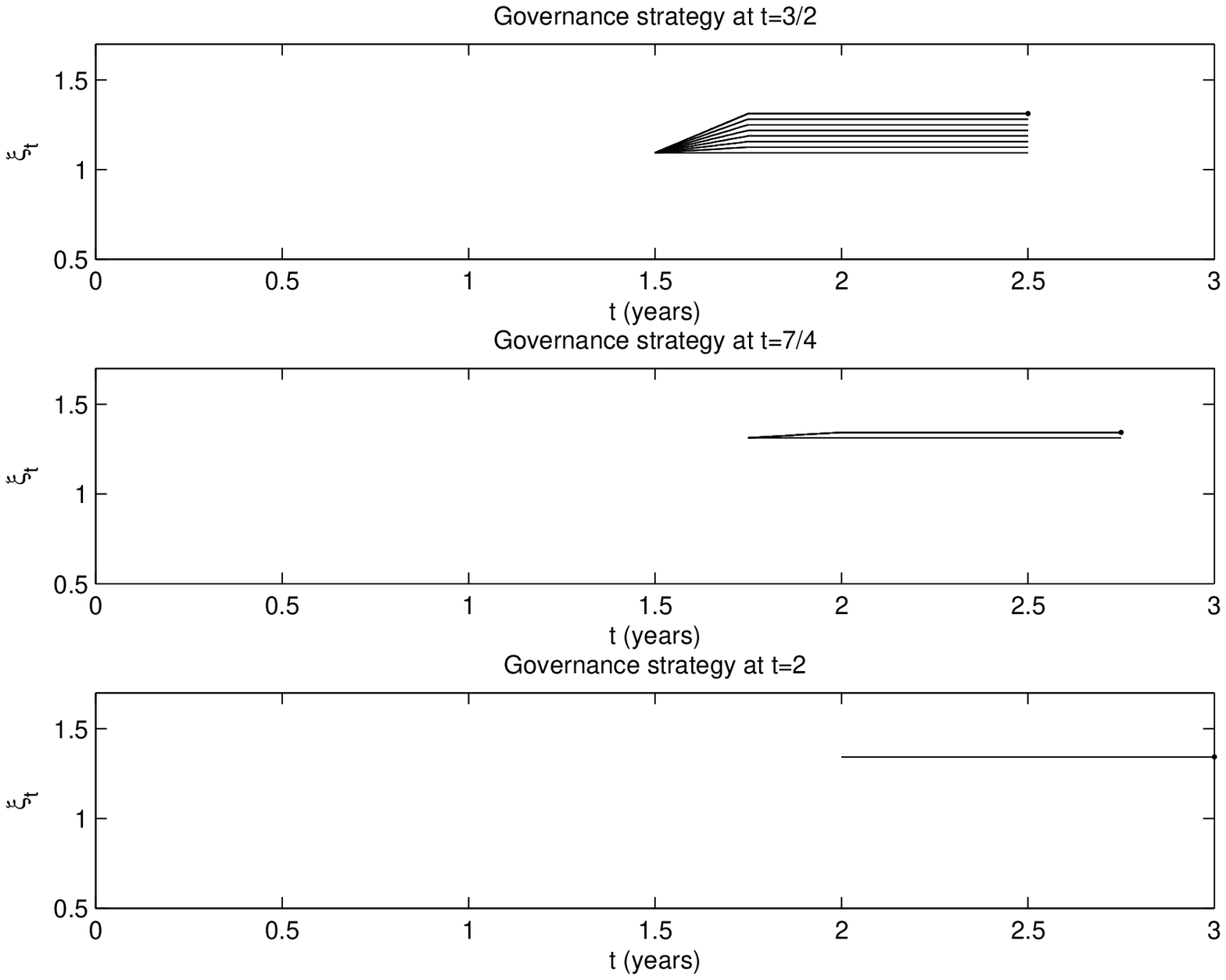}}
  \caption{\footnotesize  Numerical experiment 3:   choices of target trajectory  considered by the monetary authority at the beginning of a quarter;  the target trajectory   marked with a   dot at its endpoint  is  the final choice of  the monetary authority.}
 \label{fig18}
  \end{figure}
  \begin{figure}
 \centerline{\includegraphics[height=8cm]{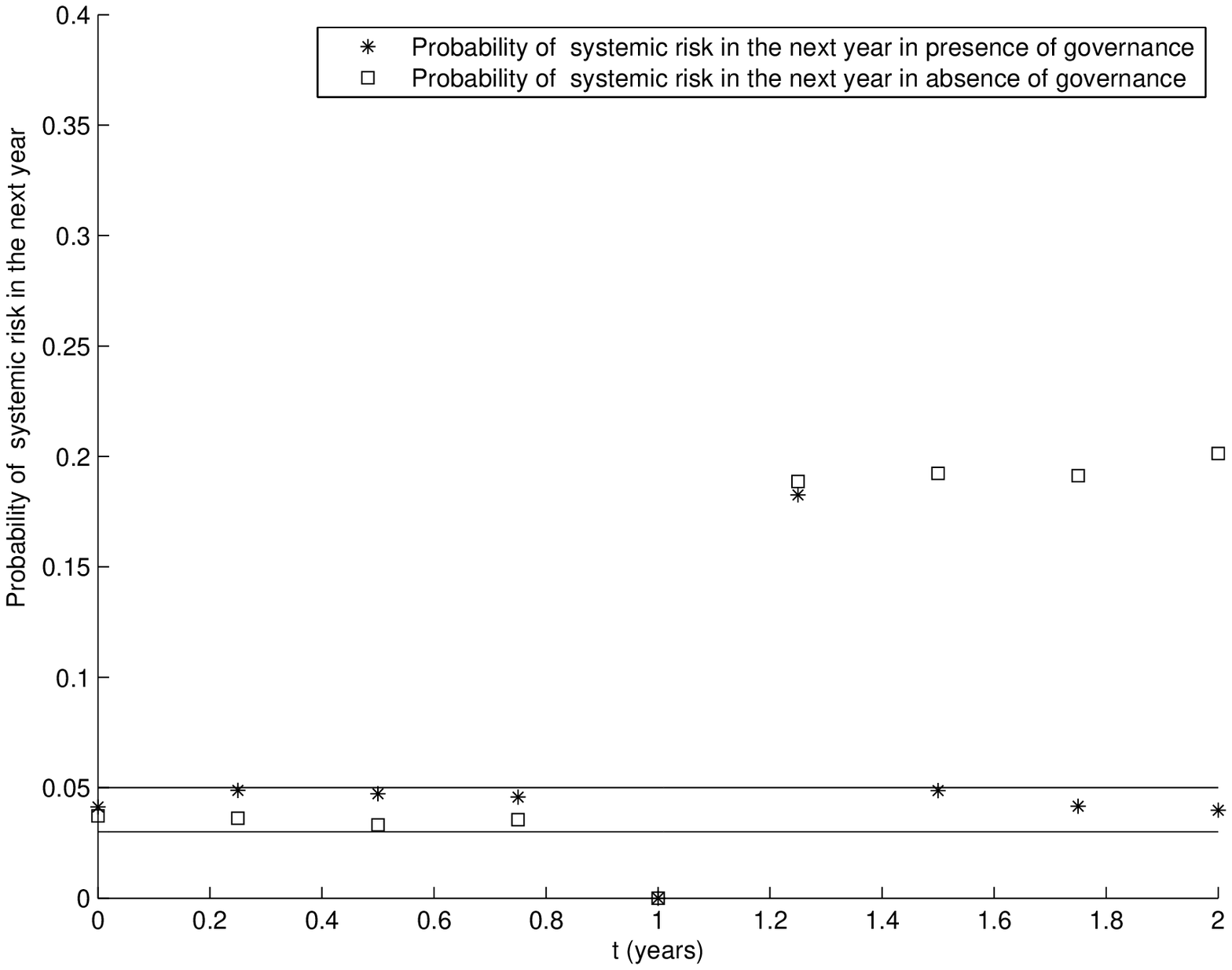}}
 \caption{\footnotesize    Numerical experiment 3: probability of   systemic risk in the next year evaluated  at  the beginning of each quarter 
  in the time interval  $ [0,T_{2}]= [0,2]$   in presence of governance ($*$) and in  absence of governance ($\square$).
}
 \label{fig19}
  \end{figure}
 %
 %
 %
  

\end{document}